\documentclass[aps,prx,reprint,amsmath,showpacs,amsfonts,amssymb,longbibliography]{revtex4-1}
\usepackage{graphicx}
\usepackage{bm}
\usepackage{color,soul}
\usepackage{MnSymbol}
\usepackage{enumerate}
\usepackage{bbold}
\usepackage{lipsum}
\usepackage{array}
\usepackage{textgreek}
\usepackage{amsmath}
\usepackage{amsfonts}
\usepackage{amssymb}
\usepackage{graphicx}
\usepackage{mathtools}
\usepackage{multirow}
\usepackage{hyperref}
\usepackage{float}
\usepackage{blindtext}
\usepackage{pifont}% http://ctan.org/pkg/pifont
\newcommand{\cmark}{\ding{51}}%
\def\la{\langle}
\def\ra{\rangle}

\def\be{\begin{equation}}
\def\ee{\end{equation}}
\newcommand{\op}[1]{{\hat #1}}
\newcommand{\dB}{\dd\op{B}}

\newcommand{\Eavg}[1]{{{\rm E}\!\left\{ #1 \right\}}}

\newcommand{\Od}{\op{d}}
\newcommand{\Odi}{\op{d}_1}
\newcommand{\Odii}{\op{d}_2}

\newcommand{\beq}{\begin{equation}}
\newcommand{\eeq}{\end{equation}}

\newcommand{\bra}[1]{\langle{#1}|}
\newcommand{\ket}[1]{|{#1}\rangle}

\newcommand{\Tr}[1]{{\rm Tr}\left\{#1\right\}}

\definecolor{nblue}{rgb}{0.06,0.3,0.73}%229 11R, 61G, 145B
\definecolor{patriarch}{rgb}{0.5, 0.0, 0.5}
\definecolor{nblack}{rgb}{0,0,0}
\definecolor{nred}{rgb}{0.9,0.1,0.1}
\definecolor{nmagenta}{rgb}{0.7,0.0,0.3}
\definecolor{newc}{rgb}{0.5,0.3,0}
\newcommand{\blu}{\color{nblack}}
\newcommand{\red}{\color{nblack}}

\newcommand{\blk}{\color{nblack}}
\newcommand{\newc}{\color{newc}}

\newcommand{\rep}[2]{\st{#1}{\newc #2}}

\newcommand{\dd}{{\rm d}}
\newcommand{\ddt}{\Delta t}
\newcommand{\dt}{\dd t}

\newcommand{\subc}{_{r}}

\usepackage[T1]{fontenc}
\DeclareFontFamily{T1}{calligra}{}
\DeclareFontShape{T1}{calligra}{m}{n}{<->s*[1.44]callig15}{}
\DeclareMathAlphabet\mathcalligra   {T1}{calligra} {m} {n}
\DeclareMathAlphabet\mathzapf       {T1}{pzc} {mb} {it}
\DeclareMathAlphabet\mathchorus     {T1}{qzc} {m} {n}
\DeclareMathAlphabet\mathrsfso      {U}{rsfso}{m}{n}

\makeatletter
\newcommand*\bigcdot{\mathpalette\bigcdot@{.5}}
\newcommand*\bigcdot@[2]{\mathbin{\vcenter{\hbox{\scalebox{#2}{$\m@th#1\bullet$}}}}}
\makeatother

\begin{document}

\title{Completely positive trace-preserving maps for higher-order unraveling\\ of Lindblad master equations}
%\title{Entanglement trajectories, distributions, and most likely paths for jointly measured qubits in remote cavities}

\author{Nattaphong Wonglakhon$^1$}
\author{Howard M. Wiseman$^1$}
\author{Areeya Chantasri$^{1,2}$}
\affiliation{$^1$Centre for Quantum Computation and Communication Technology (Australian Research Council), \\ Centre for Quantum Dynamics, Griffith University, Yuggera Country, Brisbane, Queensland 4111, Australia\\
$^2$Optical and Quantum Physics Laboratory, Department of Physics, Faculty of Science, Mahidol University, Bangkok 10400, Thailand
}

\date{\today}

\begin{abstract}
Theoretical tools used in processing continuous measurement records from real experiments to obtain quantum trajectories %or to unravel the Lindblad master equations 
can easily lead to numerical errors due to a  non-infinitesimal time resolution. In this work, we propose a systematic assessment of the accuracy of a map. We perform error analyses for diffusive quantum trajectories, based on single-time-step Kraus operators proposed in the literature, and find the orders in time increment, $\ddt$, to which such operators
satisfy the conditions for valid average quantum evolution (completely positive, convex-linear, and trace-preserving), and the orders to which they match the Lindblad solutions. 
%can satisfy \red our so-called \emph{valid average quantum evolution} conditions, including \blk  completely positive, convex-linear and trace-preserving conditions, along with reproducing the Lindblad solutions. 
%Compared with the usual approaches, e.g., using the It\^o stochastic master equation, which assumes infinitesimal time resolution, 
Given these error analyses, we propose a Kraus  operator that satisfies the valid average quantum evolution conditions and agrees with the Lindblad master equation, to second order in $\ddt$, thus surpassing all other existing approaches. 
%We construct the new map from the first principle, assuming a quantum system coupled to a bosonic bath, deriving their unitary map, and carefully choosing terms that satisfy our desired conditions. 
In order to test how well our proposed operator reproduces exact quantum trajectories, we analyze two examples of qubit measurement, where exact maps can be derived: a qubit subjected to a dispersive ($z$-basis) measurement and a fluorescence (dissipative) measurement. We show analytically that our proposed operator gives the smallest average trace distance to the exact quantum trajectories, compared to existing approaches. %Moreover, using the linear quantum trajectory technique, we can also show that our map satisfy both the complete positivity condition and the perfect Lindblad unravelling up to the second order in the time increment.
\end{abstract}
\maketitle
\date{today}

%%%%%%%%%%%%%%%%%%%%%%
% INTRODUCTION
%%%%%%%%%%%%%%%%%%%%%%%

\section{Introduction}
% [Areeya's comments on INTRO: want this paper to summarize the previous approaches, then motivate the new one with better physical motivation.

\begin{figure}[t!] \includegraphics[width=\linewidth]{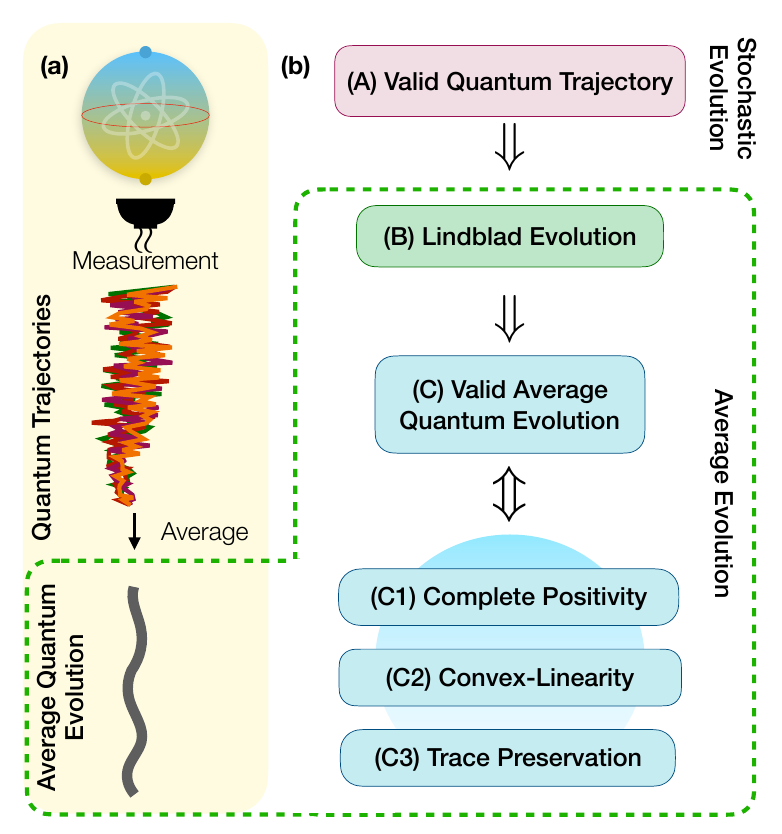} 
\caption{Diagrams show evolutions of open quantum systems and their hierarchy criteria. (a) Schematic diagrams showing a qubit measured continuously in time, resulting in quantum trajectories. (b) Hierarchy conditions (A), (B), and (C1)-(C3) and their implications are indicated by double-line arrows. If a map satisfies (A) should automatically satisfy (B) and (C1)-(C3), but not in a reverse direction (see text for more detail). Components inside the green dashed box are applied to average quantum evolutions.}
\label{fig:diagram}
\end{figure}

Open quantum system dynamics describes state evolutions of quantum systems of interest that interact with their environment (bath)~\cite{BookDavies,BookBreuer,BookNielsen}. The resulting system dynamics not only depends on the interactions with the bath, but also on whether the bath's actual states are unknown or known to observers~\cite{BookCarmichael,BookWiseman}, leading to decoherence effects or measurement backactions on the system's state, respectively. Under the strong Markov assumption~\cite{BookKampen}, if the bath's state is unknown, the system's dynamics, after tracing out the bath degree of freedom, exhibits the decoherence described by the Lindblad master equation~\cite{Lind1976,LiLi2018}. However, if the bath's state is revealed, via measurement, the system's state can be estimated \emph{conditioned} on measurement outcomes. For continuous quantum measurement, a realization of the measurement record is stochastic by nature, leading to a stochastic conditioned state dynamic referred to as a \emph{quantum trajectory} or \emph{quantum state filtering}~\cite{Bel99,BookWiseman,BookJacobsSto}. The quantum trajectory can also be considered as a result of \emph{unravelling} the Lindblad master equation, because, by averaging the conditioned dynamics over all possible measurement records, their average coincides with the solution of the master equation, see Fig.~\ref{fig:diagram}(a). \blk

To unravel the Lindblad master equation using continuous quantum measurement, many theoretical formulations have been proposed and even implemented in real experiments. The most common form of unravelling is via stochastic Schr{\"o}dinger equations (SSEs) and stochastic master equations (SMEs)~\cite{Davies1969,BookGardiner}, which can be derived from quantum (filtering) stochastic calculus~\cite{BookBelavkin,Belavkin1992,Barchielli1993}, from system-bath interactions in quantum optics~\cite{BookCarmichael,Barchielli1990,Wiseman1993,Wiseman1993-2,Plenio1998}, or from phenomenological Bayesian-type update equations~\cite{Korotkov1999,Korotkov2001}. The SSEs and SMEs have been the basic formalisms used in describing stochastic quantum evolutions~\cite{Diosi1988,Wiseman1993-2,Jacobs2006,Gambetta2008}, for qubits and other quantum systems; see Refs.~\cite{Roch2014,BarCom2017,IvaIva2020,GuiRou2022,SteDas2022} for recent work. However, such SMEs are typically written with the assumption that the noisy measurement records from weak (diffusive) measurements can be written as a sum of signal parts and Gaussian white noises.
%It\^o interpretation~ and 
This is only strictly valid for the continuous time limit, i.e., when the time increment is assumed infinitesimal ($\dt$). Unfortunately, measurement records from real experiments are obtained with finite time resolution, denoted by $\ddt$. Therefore, directly using the SMEs to process the finite-time discretized experimental data likely results in numerical errors. For example, directly integrating an SME for quantum states when the time step is not small enough can result in quantum states being non positive semi-definite or even not normalized~\cite{AmiRou2011,RalJac2011,AmiRou2014,Rouchon2015}.

There have been proposed workarounds to reduce such numerical errors by, instead of integrating SMEs, computing quantum state evolutions via $\ddt$-discretized maps, using some approximated measurement operators or maps. In the early work to verify quantum trajectories from homodyne signals in  superconducting circuits~\cite{Vijay2012,Murch2013,Weber2014,chantasri2016,Shay2016noncom}, the quantum states were computed with maps constructed from Bayesian probabilities, which could give exact quantum evolutions, but only for Hermitian measured observables. %The SMEs could also be used, but they need special care e.g. use with small resolution to process the finite-time data~\cite{Roch2014}.
For a more general treatment, a positivity preserving formulation~\cite{AmiRou2011,AmiRou2014} has also been used in processing superconducting qubit's heterodyne fluorescence measurement~\cite{Campagne-Ibarcq2016,SixCam2016,ROUCHON2022}, or even combined with a strongly convergent stochastic integration method such as the Euler-Milstein~\cite{Milstein1995,Rouchon2015}. However, as we show in this work, those maps reproduce the Lindblad evolution, and generate valid average quantum evolution, with an accuracy only to first order in $\ddt$. Even in a recently proposal by Guevara and Wiseman~\cite{Guevara2020}, in which the map generates valid average quantum evolution to second order in $\ddt$, it does not reproduce the Lindblad evolution to the same accuracy. This means that quantum trajectories generated with all the above approaches only average to the Lindblad solution to first order in $\ddt$.%~\cite{Steinbach1995}.

%Working in a finite $\ddt$ may cause errors in quantum trajectory. TERRIBLE INTRODUCTORY SENTENCE!

In this work, we present a systematic way to analyze what should be a good map in quantum trajectory theory. The diagram in Fig.~\ref{fig:diagram}(b) shows a hierarchy of conditions. The strongest one is (A) \emph{valid quantum trajectory}. This means that each trajectory generated with a finite $\ddt$ agrees, to some order in $\ddt$ (suitably defined), to the corresponding exact trajectory which would be generated with $\ddt \to \dt$, where $\dt$ is an infinitesimal time. If this is satisfied, then the average evolution will satisfy (B) \emph{Lindblad evolution}, to the same order in $\ddt$. In turn, satisfying the Lindblad evolution means the trajectories on average also satisfy (C) \emph{valid average quantum evolution}, to the same order in $\ddt$. This last can be broken into three conditions that must be satisfied: (C1) \emph{Complete positivity}, (C2) \emph{Convex-linearity}, and (C3) \emph{Trace preservation}. These ideas will be made more precise in Section~\ref{OQS}. 

None of the one-way implications in Fig.~\ref{fig:diagram}(b) work in the reverse direction. 
%The previous works have unsystematically considered (B) only in the first order in $\ddt$. 
For example, the Guevara-Wiseman map~\cite{Guevara2020}, satisfies (C) to second order in $\ddt$, but satisfies (B) only to first order. This was stated explicitly in Ref.~\cite{Guevara2020}, but other higher-order approaches have not been analyzed according to this hierarchy. In this paper, we analyze these existing map systematically and show their $\ddt$ order of error in the hierarchy (A)-(C), also using the breakdown (C1)-(C3). We show that all existing approaches satisfy (B) only at the first order in $\ddt$. This brings us to propose a higher-order measurement map, which we call the W-map,  %\hmw{WWC-map is too obviously named after ourselves. We are not allowed to do that. We could get away with W-map (or see below)} %called the WWC-operator, 
constructed from a unitary system-bath interaction expanded to fourth order in the bath operator. This W-map satisfies the conditions (B) and (C1)-(C3), all to the second order in $\ddt$.

On the level of individual trajectory, we can also check whether condition (A) is satisfied to higher order for some specific instances, where \emph{exact} (or nearly exact) finite-time solutions for quantum trajectories can be derived analytically. This is done for two qubit examples: the qubit continuous $z$-measurement, and the fluorescence homodyne measurement. We calculate how closely each of the $\ddt$-maps is to the corresponding map from the exact quantum trajectories, and average over all an ensemble of records. 
%There have been derived the \emph{exact} maps for the two qubit measurements in literatures, despite lacking of general measurement setups, for both qubit measurement setups.  The former setup have used in superconducting qubit~\cite{Murch2013} which is constructed from Bayesian technique~\cite{Korotkov1999,Korotkov2001,Korotkov2002,Ruskov2003,Gambetta2001,Gambetta2005,Tsang2009,Tsang2009-1,Tsang2009-2,Tsang2013,Kiilerich2016,Zhang2017,Six2016,Albarelli2018,Six2015}. The latter exact map is known as a non-Hermitian measurement, which is a map from the solution of solving the It\^o equation~Ref[]. We note that the two exact measurement operators can be shown to guarantee the complete positivity and are \emph{independent} of the size of the time resolution $\dt$. 
We show that our proposed W-map %\hmw{PRA will not let us use the word ``new''. So always say W-map, not new map. Or we could use N-map, which could be Nattaphong map or (secretly) new map. :-)} 
gives the best accuracy among all existing methods, although there is no difference in scaling with $\ddt$ for any of the measurement operators we consider. Achieving (A) to higher order is thus still an open problem for higher order quantum trajectory theory. \blk %We also note that, in this work, we only consider a single measurement channel. The full generalized version of the map can be too complex to be implemented in practice, therefore, the effects of Hamiltonian and extra dephasing can be added separately.

The paper is organized as the follows. In Section~\ref{OQS}, we summarize theories for open quantum systems, describing conditioned (measurement) evolution, unconditioned (Lindblad) evolution, and valid average quantum evolution, along with introducing properties (A), (B), and (C1)-(C3). Here measurement operators and their properties are defined. In Section~\ref{ExisApps}, we investigate measurement operators of existing approaches in the literature: the standard It\^o map, the Euler-Milstein approach~\cite{Rouchon2015}, and the so-called completely positive map~\cite{Guevara2020}. In Section~\ref{Construct}, we derive our W-map, from a system coupled to a bosonic (harmonic-oscillator) bath, and show that it satisfies properties (B) and (C) to second order in $\ddt$. We then consider the two qubit examples of continuous $z$-measurement and homodyne fluorescence measurement, in Section \ref{MapCompare}, showing the deviation of individual quantum trajectories from their exact counterparts for all the approaches. Section~\ref{conclusion} concludes.

%%%%%%%%%%%%%%%%%%%%%%
% OPEN QUANTUM SYSTEM AND MEASUREMENT OPERATION
%%%%%%%%%%%%%%%%%%%%%%%
\section{Open Quantum System and Measurement Operation}\label{OQS}
% from open quantum system to measurement operation
\par Let us consider the time evolution of a quantum system interacting with its environment (or bath). The system and the bath together, called a combined system, can be treated as a closed quantum system, where its evolution can be described by a unitary evolution with a specific Hamiltonian. Pure quantum states of the combined system live in a tensor product of individual Hilbert spaces, $\mathcal{H}_s\otimes \mathcal{H}_e$, where $\mathcal{H}_s$ and $\mathcal{H}_e$ are the Hilbert spaces of the system of interest and its bath, respectively. 
For applications in continuous measurement under the strong Markov assumption, we assume that the system interacts with its bath for a short interval between time $t$ and $t+\ddt$ and that the combined system at the start of any interval is in a product state $\varrho(t) =\rho(t) \otimes \ket{e_0}\bra{e_0}$, where $\ket{e_0}\bra{e_0}$ is the initial bath's state. In our work, we do not assume the time interval $\Delta t$ to be infinitesimally small in order to investigate effects from its being finite. The unitary evolution of the combined system is described by
\begin{eqnarray}\label{CombinedEvo}
\varrho(t+\ddt)&=&\hat{U}_{t+\ddt,t}\left(\rho(t) \otimes\ket{e_0}\bra{e_0} \right)\hat{U}^\dagger_{t+\ddt,t},
\end{eqnarray}
where $\hat{U}_{t+\ddt,t}$ is the unitary operator for the combined system.
After the evolution in Eq.~\eqref{CombinedEvo}, the system may become entangled with the bath. %To obtain the reduced dynamics of the (open) quantum system alone, the reduced state can be found by taking a trace on $\varrho(t+\ddt)$ over the bath's degree of freedom. %, one can trace out the bath's degrees of freedom from the combined system's dynamics. 

%In the regime where the bath's state is not observed, or equivalently no information available about the bath's state after the interaction, the reduced density matrix for the system's state can be found by taking a trace over the bath's degree of freedom, resulting in the \emph{decoherence} (non-unitary evolution) \cite{BookWiseman,LiLi2018,BookNielsen,Brune1996} on the system's state. The decoherence can be thought of as a result of noises from the bath affecting the system. However, in the case where there is an observation, i.e., the bath's state is measured by some measurement detection scheme, then the system's state should reflect the information gained from measuring the bath. We will see more discussions on the conditioned evolution in Sections \ref{SecConEvo} and Chapter~\ref{ExisApps}.%we then obtain the Lindblad master equation (unconditioned evolution).
%%%%%%%%%%%%%%
\subsection{Conditioned state evolution on\\ measurement readouts}\label{SecConEvo}
Let us consider a scenario where there is an observation, i.e., the bath's state is measured by an observer with some detection scheme. The system's state evolution should therefore reflect the information gained from the measurement readouts. To describe the bath's detection, we consider collapse of the bath's state to one of the eigenstates of the bath's measured observable. Let us define one of the eigenstates by $\ket{e\subc} \in \{ \ket{e_k}\}$, which also corresponds to a particular observed readout (`$r$' stands for `readout'). The conditioned system's state, after tracing out the bath's degrees of freedom from the combined system's state, $\varrho(t+ \ddt)$, in Eq.~\eqref{CombinedEvo}, becomes
\begin{eqnarray}\label{ConDyn}
\rho\subc(t+\ddt)&\propto&\text{Tr}_e[\ket{e\subc}\bra{e\subc}\varrho(t+\ddt)]\nonumber\\
&=&\hat{K}(r)\rho(t)\hat{K}^\dagger(r) \equiv \mathcal{J}[\hat{K}(r) ]\rho(t),
\end{eqnarray}
where a superoperator $\mathcal{J}[\op A]\bullet\equiv \op A \bullet\op A^\dagger$ represents a linear map with an operator $\op A$. Noting that the equations on the right side are not normalized. %Their first line is from the combined system state $\varrho(t+\Delta t)$ being acted by a projector, $\ket{e\subc}\bra{e\subc}$, before the trace, while, 
In the second line, we have defined a measurement operator
\begin{eqnarray}\label{MeasOperator}
\hat{K}(r)&=&\bra{e\subc}\hat{U}_{t+\ddt,t}\ket{e_0},
\end{eqnarray}
commonly known as a Kraus operator. We note that Eq.~\eqref{ConDyn}, even though not normalized, has a linear property in the quantum state and can be used in generating \emph{linear} quantum trajectories~\cite{BookWiseman,GG1994,Jacobs1998}. These play an important role in the quantum state retrodiction~\cite{Gammelmark2013}, the two-state formalism~\cite{AhaVai2002} and the quantum state smoothing~\cite{Guevara2015,chantasri2019,Guevara2020}, as will be explained in Sec.~\ref{VAQE}. 

By normalizing Eq.~\eqref{ConDyn}, we obtain a nonlinear superoperator,
 \begin{align}\label{ConEvo}
\rho\subc(t+\ddt)=  \frac{\hat{K}(r)\rho(t)\hat{K}^\dagger(r)}{{\rm Tr}[\hat{K}(r)\rho(t)\hat{K}^\dagger(r)]} \equiv  \tilde{\cal J}[{\hat K}(r)] \rho(t),
\end{align}
which ``maps'' a quantum state $\rho(t)$ at time $t$ to a new state $\rho(t+\Delta t)$ at the later time via the Kraus operator $\op K(r)$. This nonlinear map is commonly used in generating a so-called \emph{normalized} state for the trajectories.%~\cite{Guevara2015}

Properties of the ${\hat K}(r)$ operator are closely related to statistics of the measurement readout $r$. The denominator of Eq.~\eqref{ConEvo} is to be interpreted as the probability (or probability density for the case of continuous $r$) of getting the readout given that the system's state before the measurement was $\rho(t)$. That is to say, the probability density function (PDF) is
\begin{eqnarray}\label{eq-probygen}
  \wp({r} \, |\rho(t)) = {\rm Tr}[\hat{K}(r)\rho(t)\hat{K}^\dagger(r)],
\end{eqnarray}
which can be used in computing various statistical quantities such as the mean readout and its variance,
\begin{subequations}
\begin{align}
    \label{eq-mu}\mu &= \Eavg{r},\\
    \label{eq-sigma}\sigma^2 &=\Eavg{r^2} - \Eavg{r}^2,
\end{align}
\end{subequations}
where we have defined the expectation value notation, which will be used throughout the paper,
\begin{align}\label{everagedefo}
{\rm E}\{ f({r})\} \equiv \int \!\! {\rm d} r\,\,  \wp({r}\, |\rho(t)) f({r}),
\end{align}
for any function of the readout $r$. Therefore, given Eq.~\eqref{eq-probygen} and its normalization condition, the Kraus operator should satisfy a \emph{completeness relation}
\begin{align}\label{CompletenessRelation}
\int \!\!{\rm d}r \,\hat{K}^\dagger(r) \hat{K}(r) 
= \,\, \hat 1,
\end{align}
which is the identity in the system's state space. One can also get the same condition by using Eq.~\eqref{MeasOperator} and show that $\int \!\!{\rm d}r \,\bra{e_0}\hat{U}_{t+\ddt,t}^\dagger \ket{e\subc} \bra{e\subc}\hat{U}_{t+\ddt,t}\ket{e_0} = \hat 1$, where the integral over all possible readouts of the bath gives the identity in the bath's state space, i.e.,  $\int \!{\rm dr} \, \ket{e\subc} \bra{e\subc} = \hat 1_e$. This condition Eq.~\eqref{CompletenessRelation} is also called a trace-preserving condition for a map with Kraus operators.

However, the Kraus operator, $\hat K(r)$, in the form of Eq.~\eqref{MeasOperator} cannot always be derived exactly, %as one needs to know the interaction unitary operator $\op U(t+\ddt,t)$, or even if we know the interaction perfectly, the exact operator can be mathematically too complicated to obtain. 
as it is a mathematically complicated object. Therefore, one usually works with approximated versions. For describing continuous-in-time measurement, %where measurement records were obtained from small (to infinitesimal) time interval, 
the most used approximation is the expansion to first order in the time increment. This could lead to errors in the quantum state calculation, as mentioned in the introduction, when the time increment $\ddt$ of a measurement record is not really infinitesimal. Therefore, we here introduce a valid quantum trajectory condition: \\

\emph{ Condition (A) Valid quantum trajectory.} A map satisfies this condition if it generates a quantum trajectory that matches with $\rho\subc(t)$ generated from Eq.~\eqref{ConEvo}, using the (exact) Kraus operator in Eq.~\eqref{MeasOperator}. \\

We will consider this condition in detail later in Section~\ref{qubitExample}, where we introduce the two examples of a qubit system with solvable exact maps. %Note that the exact quantum trajectory would also be simulated by taking $\ddt\rightarrow\dt$.

%%%%%%%%%%%%%%
\subsection{Unconditioned state evolution}\label{SubLin}
Continuing from Eq.~\eqref{CombinedEvo}, if the bath's state is not observed, the reduced density matrix for the system's state can be found by taking a trace over the bath, resulting in the decoherence (non-unitary evolution) on the system's state. From Eq.~\eqref{CombinedEvo}, we can write an unconditioned system's state as 
\begin{eqnarray}\label{traceCombine}
\rho(t+\ddt)&=&{\rm Tr}_e\left[ \hat{U}_{t+\ddt,t}(\rho(t)\otimes\ket{e_0}\bra{e_0})\hat{U}^\dagger_{t+\ddt,t} \right],
\end{eqnarray}
where the trace ${\rm Tr}_e[\cdots] \equiv \sum_k \la e_k| \cdots | e_k \ra$ is defined for the eigenbases of the bath's Hilbert space. Under the Markov assumption, one finds that the reduced system's dynamics Eq.~\eqref{traceCombine} can be reformulated as~\cite{Lind1976,LiLi2018},
\begin{eqnarray} \label{stateupdate0}
\rho(t+\ddt)&=&e^{\ddt \mathcal{L}\bullet }\rho(t),
\end{eqnarray}
where  
\begin{align}\label{eq-Lop}
\mathcal{L}\bullet=-i[\hat{H}, \bullet]+\sum_{j=1}^N\mathcal{D}[\hat{c}_j]\bullet,
\end{align}
is the Lindblad superoperator. Here, the operator $\hat{H}$ is a Hermitian operator describing unitary dynamics of the system's state, the superoperator $\mathcal{D}$ is defined as $\mathcal{D}[\hat c]\bullet=\hat c \bullet \hat c^\dagger-\frac{1}{2}\big\{\hat c^\dagger\hat c,\bullet\}$ describing the non-unitary dynamics, and $\hat{c}_j$ are the Lindblad operators describing $N$ couplings between the system and the bath. We note that we have taken $\hbar=1$ throughout this paper.

In the time-continuum limit, as usually assumed in most works in the literature, $\ddt \rightarrow \dd t$, where $\dt$ is an infinitesimal time, Eq.~\eqref{stateupdate0} becomes the Lindblad master equation in differential form,
\begin{eqnarray}\label{LinME} 
\partial_t\rho(t)=\mathcal{L}\rho(t).
\end{eqnarray}
However, in experiments~\cite{Murch2013,Weber2014,chantasri2016,Shay2016noncom}, $\ddt$ cannot be infinitesimal, the exact incremental evolution Eq.~\eqref{stateupdate0} has to be used. If the exponential cannot be solved exactly, it can be solved by expanding to some orders in the time increment $\ddt$. The higher the order of expansion, the higher the accuracy of the state calculation.

Let us consider the case with a single Lindblad operator $\hat{c}$ and no unitary dynamics, $\hat{H}=0$, in order to focus on the dynamics of a single decoherence channel. The expansion to first order in $\ddt$ is trivial:
\begin{align}\label{SMEav}
    \rho(t+\ddt)&=\left(\op 1+\ddt \mathcal{L}\right)\rho(t)=\rho(t)+\ddt\mathcal{D}[\op c]\rho(t).
\end{align} 
Beyond that, it has been shown by Steinbach~\emph{et al.}~\cite{Steinbach1995} that the Taylor expansion of Eq.~\eqref{stateupdate0} to second order in $\ddt$ gives
\begin{eqnarray}\label{stateupdate}
	\rho(t+\ddt)&=& \bigg(\hat{1}+\ddt\mathcal{L}+\tfrac{1}{2}\ddt^2\mathcal{L}^2\bigg)\rho(t),\\
&=&\rho(t)+\ddt \mathcal{D}[\hat{c}]\rho(t)+\tfrac{1}{2}\ddt^2\mathcal{D}^2[\hat{c}]\rho(t),\nonumber
\end{eqnarray}
where the double Lindblad operator can be expressed explicitly as
\begin{align}\label{LindHigh}
\mathcal{D}^2[\hat{c}]\rho(t) = &+\tfrac{1}{2}\hat{c}^\dagger\hat{c}\rho(t)\hat{c}^\dagger\hat{c} +\hat{c}^2\rho(t)(\hat{c}^\dagger)^2 \nonumber \\
& + \tfrac{1}{4}\big[\rho(t)(\hat{c}^\dagger\hat{c})^2+(\hat{c}^\dagger\hat{c})^2\rho(t)\big] \nonumber \\
&-\tfrac{1}{2}\big[\hat{c}\rho(t)(\hat{c}^\dagger)^2\hat{c}+\hat{c}^\dagger\hat{c}^2\rho(t)\hat{c}^\dagger\big] \nonumber \\
& -\tfrac{1}{2}\big[\hat{c}\rho(t)\hat{c}^\dagger\hat{c}\hat{c}^\dagger+\hat{c}\hat{c}^\dagger\hat{c}\rho(t)\hat{c}^\dagger\big].
\end{align}
Moreover, the conditioned evolution Eq.~\eqref{ConEvo} can also be considered as an unravelled trajectory of the unconditioned state (Lindblad) evolution, Eq.~\eqref{traceCombine}, where we can get back the dynamics \eqref{stateupdate0} by integrating the conditioned state over all possible readouts with their probability weights. That is to say,
\begin{align}\label{eq-Lindblad}
\rho(t+\ddt)&=\Eavg{\tilde{\cal J}[\hat{K}(r)] \rho(t)}\nonumber \\ 
& = \int \!\! {\rm dr}\,\, \wp(r|\rho(t)) \tilde{\cal J}[\hat{K}(r)] \rho(t) \nonumber\\
%&=\sum\subc \wp({\rm r} \, |\rho(t))\bar{\cal J}[\hat K\subc] \rho(t)\\ \nonumber
& = \int \!\! {\rm dr}\,\, \hat{K}(r)\rho(t)\hat{K}^\dagger(r)= e^{\ddt \mathcal{L}\bullet }\rho(t),
\end{align}
using the definition of $\text{E}\{\bullet\}$ in Eq.~\eqref{everagedefo}, gives the Lindbad evolution, where the superoperator ${\cal L}$ is as defined in Eq.~\eqref{eq-Lop}.
This leads to defining our second condition:\\

\emph{Condition (B) Lindblad Evolution.} A map satisfies this condition if it generates quantum trajectories that average to the Lindblad equation as in Eq.~\eqref{eq-Lindblad}.\\

As we consider approximated Kraus operators in this work, we use the second-order expansion of the Lindblad evolution in Eq.~\eqref{stateupdate} for evaluating the condition (B) for high-order accuracy Kraus operators. This is analyzed in detail in Section \ref{ExisApps} and onward.

\subsection{Valid average quantum evolution}\label{VAQE}

Apart from the strong conditions (A) and (B), we here introduce weaker conditions that only require a map to generate a \emph{valid} quantum evolution on average. That is, evolution that is completely positive (C1), convex-linear (C2) and trace preserving (C3). Before we dive into the definitions of these conditions, let us consider various approaches in the literature on how quantum trajectories are generated and averaged. %the readout PDF, $\wp(r|\rho(t))$, as shown in Eq.~\eqref{eq-Lindblad}. %However, since we work on the approximated Kraus operators, the readout PDF may affect the \emph{valid average quantum evolution} (condition C), which will be explain later in this section. 

The first straightforward way is to generate normalized trajectories using the readout PDF $\wp(r|\rho(t))$, as shown in Eq.~\eqref{eq-Lindblad}, where the PDF is derived from Kraus operators (either exact or approximated ones),
\begin{align}\label{wpKraus}
    \wp_\text{k}(r|\rho(t))=\text{Tr}[\hat{K}(r)\rho(t)\hat{K}^\dagger(r)],
\end{align}
with the subscript `k' standing for `Kraus'. Therefore, if this PDF is used in generating trajectories, then their average quantum state evolution is given by:
\begin{align}\label{MethodI}
        \rho(t+\ddt)&=\int \dd r \ \wp_\text{k}(r|\rho(t))\frac{\op K(r)\rho(t)\op K^\dagger(r)}{\text{Tr}[\op K(r)\rho(t)\op K^\dagger(r)]},\nonumber\\
        &=\int \dd r \op K(r) \rho(t) \op K^\dagger(r),
\end{align} 
where the trace norm of the state exactly cancels the weight PDF. We call this as \hypertarget{method1}{\emph{Method I}}. %This method always ensures convex linearity. 
There are some interesting points to note here. If somehow an exact Kraus operator were available, but resulted in a complicate PDF function in Eq.~\eqref{wpKraus}, then it is usually not convenient to use the operator in numerically generating quantum trajectories. One can then approximate the Kraus operator as $\hat K(r)  = \sqrt{\wp_{\rm ost}}\hat M(r)$, with a simple ostensible probability $\wp_{\rm ost}$, then compute the average evolution using the second line of Eq.~\eqref{MethodI}. With this latter technique, the average can be numerically obtained from unnormalized (linear) states randomly generated with the ostensible probability, instead of the correct PDF, $\wp_\text{k}(r|\rho(t))$. This approximate technique has been used in various stochastic simulations~ \cite{Gambetta2005, Guevara2015,chantasri2019,Guevara2020, ChaGue2021}, especially in quantum state smoothing, when a large ensemble of trajectories is required.

The second way is to use a simple \emph{guessed} readout PDF, $\wp_\text{g}(r|\rho(t))$, where `g' stands for `guess' to compute the average quantum evolution. For instance in quantum optics \cite{BookWiseman}, the distribution of a homodyne detection readout is often approximated as Gaussian. The average evolution is thus given by: 
    \begin{align}\label{MethodII}
        \rho(t+\ddt)&=\int \dd r \ \wp_\text{g}(r|\rho(t)) \frac{\op K(r) \rho(t) \op K^\dagger(r)}{\text{Tr}[\op K(r) \rho(t) \op K^\dagger(r)]},
\end{align} 
and this method is called \hypertarget{method2}{\emph{Method II}}. %This method is always trace-preserving.
This method, although not quite as accurate as \hyperlink{method1}{Method I}, has advantage in numerical simulation of quantum trajectories, because a random generator with Gaussian distribution (normal distribution) is native in most programming languages.

%, where the complete positivity is always held. 
Another common approach to compute the average quantum evolution is via an SME. By naively generating states in time steps of size $\ddt$, this yields 
\begin{align}
\rho(t+\ddt)&=\rho(t)+\ddt\mathcal{D}[\op c]\rho(t)+\Delta W\mathcal{H}[\op c]\rho(t),
\end{align}
where $\mathcal{H}[\op c]\bullet \equiv \op c\bullet+\bullet\op c^\dagger-\text{Tr}[\op c\bullet +\bullet\op c^\dagger]\bullet$ and $\Delta W$ is the Wiener increment with zero mean and $\ddt$ variance~\cite{BookGardiner}. Averaging over the realizations of the Wiener increment gives $\rho(t+\ddt)=\rho(t)+\ddt \mathcal{D}[\op c]\rho(t)$, to first-order in $\ddt$,  as given earlier in Eq.~\eqref{SMEav}. %However, for time-discret SME, the completely positivity condition is usually violated. 
Although the SMEs are not employed for unraveling quantum evolution in this work, we present their properties to provide a comprehensive comparison with other methods.

\begin{table}[t!]
 \renewcommand{\arraystretch}{1.3}
\begin{tabular}{|c | c | c| c|}
	\cline{2-4}
    \multicolumn{1}{c|}{} & \multicolumn{3}{c|}{Satisfaction}  \\ [1.5ex] 
    \hline 
	Conditions & \begin{tabular}{@{}c@{}} (C1) Complete   \\ positivity \end{tabular} & \begin{tabular}{@{}c@{}} (C2) Convex-   \\ linearity \end{tabular} & \begin{tabular}{@{}c@{}} (C3) Trace   \\ preservation \end{tabular}\\  [1.75ex] 
	\hline
	 \hyperlink{method1}{Method I}& \cmark & \cmark & \begin{tabular}{@{}c@{}} iff  \\ $\int \dd r \op K^\dagger_r\op K_r=\op 1$  \end{tabular} \\ [1.75ex] 
	\hline
    \hyperlink{method2}{Method II} & \cmark & \begin{tabular}{@{}c@{}} iff $\wp_\text{g}(r|\rho(t))\propto$\\ $\text{Tr}[\op K_r\rho(t)\op K^\dagger_r]$\end{tabular} & \cmark  \\ [1.5ex] 
	\hline
   Av. SME & \begin{tabular}{@{}c@{}} Errors  \\ \footnotesize  at $\mathcal{O}(\ddt^2)$ \end{tabular}& \cmark & \cmark  \\ [1.75ex] 
	\hline
\end{tabular}
\caption{Summary of conditions for each method of valid average quantum evolution to satisfy (\cmark) the three conditions (C1)-(C3) in Fig.~\ref{fig:diagram}(b). %Method I is computed via Eq.~\eqref{MethodI}, while Method II via Eq.~\eqref{MethodII}. 
We use ``Av. SMEs'' as a short hand for the averaged SME. Note that we have simplified the notation by using $\op K_r \equiv \op K(r)$.}
\label{Summary-Error}
\end{table} 

Given all these different maps and methods to obtain average quantum evolutions, it is natural to ask which method is better when $\ddt$ is finite, when the Kraus operator is typically no longer exactly as Eq.~\eqref{MeasOperator} (or the SME is only first order in $\ddt$). To address this question, a systematic criterion is needed to evaluate the validity of maps for the average evolution. We consider the three conditions (C1)-(C3) for valid average quantum evolution, as illustrated in the hierarchy diagram in Fig.~\ref{fig:diagram}(b). The three conditions, as summarized in Table~\ref{Summary-Error}, are stated as follows:\\

\emph{Condition (C1) Complete Positivity.} A map is completely positive when its acting on part of a bipartite quantum state is positive. That is: it maps a positive state to a positive state. \\

This condition is always satisfied for a map with Kraus operators, for instance, \hyperlink{method1}{Method I} and \hyperlink{method2}{Method II}, ensure complete positivity by construction. By contrast, the averaged SME in Eq.~\eqref{SMEav} may violate this condition beyond first order in \(\ddt\)  (see Appendix~\ref{CPofSME} for a detailed calculation).\\

\emph{Condition (C2) Convex-linearity.} A map is convex-linear if a quantum state $\rho$ is mapped to a quantum state which is convex-linear in $\rho$. That is: a weighted mixture of two states is mapped to the same-weight mixture of the individually mapped states.\\

\hyperlink{method1}{Method I} and the averaged SME always satisfy this condition. That is because, for \hyperlink{method1}{Method I}, the normalization factor is cancelled as in Eq.~\eqref{MethodI}, leaving only the linear terms in $\rho$. For the averaged SME, the nonlinear term, ${\cal H}[\hat c]$, are also averaged out.  Interestingly, \hyperlink{method2}{Method II} fulfills (C2) if and only if the guessed PDF exactly satisfies $\wp_\text{g}(r|\rho(t)) \propto \text{Tr}[\hat{K}(r)\rho(t)\hat{K}^\dagger(r)]$, thus canceling the norm denominator in Eq.~\eqref{MethodII}.\\

\emph{Condition (C3) Trace Preservation.} A map is trace preserving if the trace of the mapped state is the same as that of the initial state.\\

\hyperlink{method2}{Method II} 
and the averaged SME always satisfy this condition. This is because, for \hyperlink{method2}{Method II} in Eq.~\eqref{MethodII}, the map is always normalized by its trace. For the trace of the averaged SME, it is always one as ${\rm Tr}[{\cal D}[\hat c]\rho]$ vanishes. However, for \hyperlink{method1}{Method I}, it satisfies this condition if and only if the completeness relation Eq.~\eqref{CompletenessRelation} holds exactly.

It is notable that the three conditions are independent as any one can fail for some method. In the next section, we take the reader through different approximations of Kraus operators that have been proposed in the literature. We will first examine \hyperlink{method1}{Method I} and \hyperlink{method2}{II} to explore the Lindblad condition (B) at high orders in $\ddt$. The average evolution, as depicted in Table~\ref{Summary-Error}, can be then used to assess the errors of conditions (C1)-(C3), depending on the method employed. For example, \hyperlink{method1}{Method I}, only the (C3) condition requires verification, while for \hyperlink{method2}{Method II}, only (C2) needs to be assessed. Finally, we investigate the validation of the quantum trajectory condition (A) by utilizing existing exact maps for the two qubit examples, which will be elaborated on in Section~\ref{MapCompare}.

%%%%%%%%%%%%%%%%%%%%%%
% EXISTING APPROACHES
%%%%%%%%%%%%%%%%%%%%%%%
\section{measurement operators from Existing Approaches}\label{ExisApps}

% what are we trying to explain here? we want to say that for diffusive measurement, how one can compute the evolution of stochastic process. From Kraus operator, then what are possible measurement that can be done, we are interested in diffusive type measurement which leads to Ito SME stochastic process. the measurement was observed in experiment, however with many different approaches used there.
%As mentioned in the previous section, if one knows the system-bath unitary operator, $\hat U_{t+\ddt, t}$, one can obtain the operator $\hat K\subc$, describing how the system's state change conditioned on the measurement result, following Eq.~\eqref{MeasOperator}. %Otherwise, one can approximate the operator from a similar phenomena of the state dynamics. There have been various approaches in constructing the measurement operator.

%Therefore, the treatment of stochastic dynamics of quantum state in this case can range from applying classical-like stochastic differential equations~\cite{BookBelavkin,Belavkin1992}, to phenomenological method~\cite{Gambetta2001,Gambetta2005,Tsang2009,Tsang2009-1,Tsang2009-2,Tsang2013,Kiilerich2016,Zhang2017,Six2016,Albarelli2018,Six2015}, where the former is derived for any arbitrary quantum experiment process, while the latter is restricted to use for some particular settings which will be discussed in Chapter~\ref{qubitExample}. 
In this section, we review the existing approaches proposed in the literature to show their satisfaction of conditions (C1)-(C3) at high-order in $\ddt$ for \hyperlink{method1}{Method I} and \hyperlink{method2}{II}. Their resulting average evolutions are also checked with the Lindblad evolution shown in Eq.~\eqref{eq-Lindblad}, expanded up to the order of $\ddt^2$. We focus particularly on the homodyne detection, as an example of the diffusive-type measurements, which has been used in quantum optics~\cite{BookCarmichael,Barchielli1990,Wiseman1993,Wiseman1993-2,Plenio1998} and circuit quantum electrodynamics~\cite{Murch2013,Weber2014,chantasri2016,Shay2016noncom} experiments. Generalization to other diffusive measurements such as the heterodyne is quite straightforward. The conditioned state dynamics from these measurements resemble diffusive stochastic processes, as the noise in the records are Gaussian white noises in the continuous-in-time limit. 

In the following subsections, we start with constructing a general form of diffusive-type measurement operators, and then investigating an operator that leads to the It\^o stochastic master equations (SMEs), the adapted version of the conventional (first-order) It\^o approach by Rouchon and Ralph~\cite{Rouchon2015} using the Euler-Milstein method, and lastly the so-called completely-positive trajectory method proposed by Guevara and Wiseman~\cite{Guevara2020}. %The first approach is constructed based on a mathematical idealization of white noises, therefore it suffers to give accurate state dynamics when used in processing measurement signals from experiments, because the time increment $\ddt$ is no longer infinitesimal. The last two approaches were then proposed to overcome the limitations. 
%As we mentioned in Section \ref{SecConEvo}, the measurement operator should satisfy the complete positivity and should lead to the unconditioned state described by the Lindblad master equation as in Eq.~\eqref{eq-Lindblad}, after averaging over all possible measurement results. 
For simplicity, we will again only consider the situation of a single Lindblad operator $\hat{c}$, and take the system's interaction-frame Hamiltonian to be $\hat{H}=0$. %The generalization to multiple channels can be found in our follow-up work~\cite{WWC2023}.

%%%%%%%%%%%%%%

\subsection{Constructing measurement operators for diffusive measurement}\label{sec-constructU}

%However, one alternative way to derive the It\^o map is via a quantum-optics approach. 

%Let us start with constructing a general form of measurement operators following Eq.~\eqref{MeasOperator}, for a quantum system undergoing a time-continuous homodyne measurement. Such the measurement can be done by first assuming the quantum system (e.g., a qubit) coupled to a Markovian infinite-mode bosonic field, with a free Hamiltonian $\hat H_{\rm F} = \sum_k \omega_k \hat b_k^\dagger \hat b_k$, where $\hat b_k$ and $\hat b_k^\dagger$ are the annihilation and creation operators for the $k$-mode field. The quantum system of interest is coupled to the bath via its Lindblad operator $\hat c$ through a dipole interaction, giving its interaction Hamiltonian $\hat V =  \sum_k g_k ( \hat b_k +  \hat b_k^\dagger)(\hat c + \hat c^\dagger)$, with the real-value coupling strength $g_k$. Applying the interaction frame and the rotating-wave approximation \cite{Wiseman1993-2,BookWiseman}, the evolution of the system-bath's state, for an infinitesimal time $\dt$, can be described by a coupling unitary operator,

Let us start with constructing a general form of measurement operators following Eq.~\eqref{MeasOperator}, for a quantum system undergoing a time-continuous homodyne measurement. As is standard, we assume the quantum system (e.g., a qubit) is coupled to a Markovian bosonic field.  The interaction Hamiltonian in the rotating-wave approximation is given by $\hat V_t =-i[\hat c\dB^\dagger_t - \hat c^\dagger\dB_t]$~\cite{Wiseman1993-2,BookWiseman,Wiseman1998,Wiseman2004}, where $\dB_t$ is an infinitesimal operator for the bath excitation. The evolution of the system-bath's state, for an infinitesimal time $\dt$, is then described by a coupling unitary operator,
\begin{align}\label{eq-uopint}
\op{U}_{t+\dt,t} = \exp\left[ \op{c} \, \dB^\dagger_t  - \op{c}^\dagger \dB_t\right].
\end{align}
The operator $\dB_t$ has the following commutator relation: 
\begin{align}\label{eq-combb}
[ \dB_t, \dB^\dagger_t] = \dt,
\end{align}
whence one can see that $\dB_t \dB_t^\dagger = 
\mathcal{O}(\dt)$.
This is consistent with defining annihilation and creation operators on a piece of field of duration $\dt$ by $\dB_t\ket{n}=\sqrt{\dt n}\ket{n-1}$ and $\dB_t^\dagger\ket{n}=\sqrt{\dt (n+1)}\ket{n+1}$, respectively. The commutator relation in Eq.~\eqref{eq-combb} will be useful in keeping orders of the expansion later.  

Now that we have the general form of the unitary operator in the interaction frame, we can calculate the system's evolution conditioned on the bath state being measured. We assume the bath initial state is in a vacuum state, i.e., $\dB_t|0\ra = 0$ for all $t$. After the interaction with the system via the unitary operator Eq.~\eqref{eq-uopint}, the bath's state is then observed via a homodyne measurement. The homodyne measurement gives information about a bath's quadrature denoted by $\hat Q_t = \dB_t + \dB_t^\dagger$. Therefore, given a measurement readout $y_s$ collected during an instantaneous time between $t=s$ and $t=s+\dt$, the homodyne measurement projects the bath's state onto an eigenstate of the quadrature, i.e, $\hat Q_s \ket{y_s} = y_s \ket{y_s}$. This eigenstate is simply the spatial coordinate of a Harmonic oscillator. One can write a Fock state $\ket{n}$ in this basis as a normalized wavefunction 
\be \label{eq-Hermite}
\la y_s | n \ra = \frac{(\alpha/\pi)^{1/4}}{\sqrt{2^n n!}} \exp(-\alpha y_s^2/2) H_n(\sqrt{\alpha} \, y_s), 
\ee 
where $\alpha = \dt/2$ and $H_n$ are the Hermite polynomials for non-negative integers $n$.

Following the definition of the Kraus operator in Eq.~\eqref{MeasOperator} and the unitary operator in Eq.~\eqref{eq-uopint}, the form of measurement operators is given by
\begin{align}\label{eq-extkraus}
\hat{K}(y_s)&= \bra{y_s}\exp\left[ \op{c} \, \dB_t^\dagger  - \op{c}^\dagger \dB_t\right]\ket{0},
\end{align}
which is not simple to compute exactly. From the commutation relation in Eq.~\eqref{eq-combb}, we know that $\dB_t \dB_t^\dagger = 
\mathcal{O}(\dt)$ in Eq.~\ref{eq-extkraus}. Therefore, one can expand the exponential function and keep terms only to some orders in $\dB$ and $ \dB_t^\dagger$. For continuous measurement, we must at least keep terms up to first order in $\dt$, which are terms with $\dB_t^2$, $\dB_t\dB_t^\dagger$, and $(\dB_t^\dagger)^2$. Let us define a big-$\mathcal{O}$ notation for the bath operator, $\mathcal{O}[|\dB_t|^n]$, for any polynomial in $\dB_t$ and $\dB_t^\dagger$ of degree up to $n$.
%\begin{align}
%    &{\cal O}[(\dB,\dB^\dagger)^n] \nonumber \\
%    &\equiv {\cal O}(\dB^n, \dB^{n-1}\dB^\dagger, \dB^{n-2}(\dB^\dagger)^2, ..., (\dB^\dagger)^n).
%\end{align}
For $n=2$, the expansion then gives
\begin{align}\label{eq-U1} 
\hat{K}(y_s) &= \bra{y_s}  \op{1} + \op{c} \, \dB_t^\dagger - \tfrac{1}{2} \op{c}^\dagger \op{c} \, \dB_t\dB_t^\dagger + \tfrac{1}{2} \op{c}^2 (\dB_t^\dagger)^2\nonumber\\
 &+ {\cal O}[|\dB_t|^3]|0 \ra, \nonumber\\
&= \bra{y_s} 0\ra \big[\hat{1}-\tfrac{1}{2}\hat{c}^\dagger\hat{c}\dt+\hat{c}y_s\dt + \tfrac{1}{2} \hat{c}^2 (y_s^2\dt^2-\dt),\nonumber\\
&+{\cal O} (\dt^{3/2}) \big]\nonumber\\
&\equiv \op K_1(y_s)+{\cal O} (\dt^{3/2}),
\end{align}
where we have already used the annihilation and creation properties of the bath operator $\dB_t$, and the wavefunction in Eq.~\eqref{eq-Hermite}. Note that $\op K_1(y_s)$ is defined to represent terms with the dimension of the first order in $\ddt$. In Eq.~\eqref{eq-U1},
we have also separated out a common term $\bra{y_s} 0  \ra$ so that the measurement operator is of the form 
\begin{align}
\hat{K}(y_s) &= \sqrt{\wp_{\rm ost}(y_s)} \hat{M}(y_s),
\end{align}
where we have defined an \emph{ostensible probability}
\begin{align}\label{ostprob-dt}
\wp_{\rm ost}(y_s) = |\bra{y_s} 0  \ra|^2 = \left(\frac{\dt}{2\pi}\right)^{1/2}\!\! \exp(-y_s^2\dt/2),
\end{align}
and an \emph{unnormalized measurement operator} 
\begin{align}\label{eq:measopgen}
\hat M(y_s) = \hat{1}-\tfrac{1}{2}\hat{c}^\dagger\hat{c}\dt+\hat{c}y_s\dt +\tfrac{1}{2} \hat{c}^2 (y_s^2\dt^2-\dt)+ {\cal O} (\dt^{3/2}),
\end{align}
which can be used in place of $\hat K(y_s)$, especially when the factor $\sqrt{\wp_{\rm ost}(y_s)}$ is not needed, such as in the normalized map Eq.~\eqref{ConEvo}.

We note that the discussion in this section only applies when $\dt$ is an infinitesimal time increment, where the measurement result $y_s$ is acquired from the detection during the time between $t = s$ and $t = s+ \dt$. However, as we mentioned in the introduction, it is impossible to implement an infinitesimal time resolution $\dt$ in a real experiment and one instead needs to consider a finite time resolution $\ddt$. This brings us to defining a coarse-grained measurement record, where the measurement result are obtained from a system with finite correlation time or a finite-bandwidth detector. If $\ddt$ is chosen as the integration time, then the coarse-grained signal is typically of the form,
\begin{align}\label{coarse-grain-weight}
Y_t&=\frac{1}{\ddt}\int_{t}^{t+\ddt}\!\! \dd s \, y_s,
\end{align}
which is a time average of the infinitesimal record $y_s$ over the time $\ddt$. For example, in superconducting qubit experiments~\cite{Roch2014,Vijay2012,Murch2013}, the integration time is carefully chosen to match cavity's decay rates (such that the Markovian assumption is still approximately valid), but still short enough compared to other time scales of the systems. %with a constant weight $1/\ddt$.

\subsection{Conventional It\^o approach}\label{ItoApp}

We first consider the conventional approach used in constructing It\^o (diffusive) stochastic differential equation (SDE) for quantum trajectories~\cite{BookBelavkin,Barchielli1990,Belavkin1992,Barchielli1993,Gardiner1992}. The treatment of the It\^o SDE for the quantum optics setting can be found in Carmichael's work~\cite{BookCarmichael}, with the derivation similar to that in the previous subsection, but with the It\^o rule, i.e., approximating that $y_s^2 \dt^2 \approx \dt$ being applied for an infinitesimal measurement result $y_s$ during an infinitesimal time $\dt$. Therefore, the fourth term in Eq.~\eqref{eq:measopgen} vanishes, leading to an operator, $\hat M(y_s) = \hat{1}-\tfrac{1}{2}\hat{c}^\dagger\hat{c}\dt+\hat{c}y_s\dt $. For a finite time increment, one can simply replace $\dt$ with $\ddt$ and the record $y_s$ with a coarse-grained record $Y_t$. However, one can also systematically derive what we call the It\^o measurement operator for a finite time increment $\ddt$. Let us discretize time to $m=\ddt/\dd t$ infinitesimal time. We can compute a product of multiple $\op M (y_s)$ defined for measurement records: $\{ y_s : s \in \{t, t+\dt, ..., t+(m-1)\dt\}$ and take the mean square limit to get (see Appendix~\ref{app-itomap} for the full derivation),
\begin{eqnarray}\label{MIto}
 \hat{M}_\text{I}(Y_t)= \hat{1}-\tfrac{1}{2}\hat{c}^\dagger\hat{c}\ddt+\hat{c}Y_t\ddt,
\end{eqnarray}
where the subscript `I' denotes `It\^o' and a normalized factor is found to be $\sqrt{\wp_{\rm ost}(Y_t)}$ given a new ostensible probability 
\begin{align}\label{Xost}
\wp_{\rm ost}(Y_t) &= \left(\frac{\ddt}{2\pi}\right)^{1/2}\!\! \exp(-Y_t^2\ddt/2).
\end{align}
This It\^o measurement operator is the most used approach in the literature, especially in theoretical work related to continuous quantum measurement. However, as we show in the following, whenever $\ddt$ is finite, this map satisfies the conditions (C1)-(C3) only to the first order in $\ddt$. Errors of  order $\ddt^2$ turn out to depend on the methods (\hyperlink{method1}{Method I} and \hyperlink{method2}{II}) we use to calculate the average, as we now show.

\subsubsection{It\^o approach Method I: Linear}\label{ItotAVLinear}

From Table~\ref{Summary-Error}, we only need to check the conditions (B) and (C3), as (C1) and (C2) are already satisfied. Given the measurement operator and its ostensible probability in Eqs.~\eqref{MIto} and \eqref{Xost}, we obtain the Kraus operator for the It\^o approach
\begin{align}
    \hat K_{\rm I}(Y_t) = \sqrt{\wp_{\rm ost}(Y_t)} \hat M_{\rm I}(Y_t),
\end{align}
which can be used to compute a readout's PDF following the definition in Eq.~\eqref{eq-probygen},
\begin{align}\label{actualprobIto}
\wp_{\rm I, k}(Y_t|\rho(t)) =\wp_{\rm ost}(Y_t){\rm Tr}\left\{{\cal J}[\op{M}_{\rm I}(Y_t)] \rho(t)\right\},
\end{align}
where the superoperator ${\cal J}[\bullet]$ was defined in Eq.~\eqref{ConDyn} and $\wp_{\rm I, k}$ is the PDF derived from the Kraus operator. One can use this PDF to compute the average dynamics following the third line of Eq.~\eqref{eq-Lindblad}. By substituting $\hat M_{\rm I}(Y_t)$ from Eq.~\eqref{MIto}, expanding terms, and integrating over $Y_t$ with the ostensible probability $\wp_\text{ost}(Y_t)$, keeping only terms up to ${\cal O}(\ddt^2)$, we obtain
\begin{align}\label{eq-aveito1}
	\rho(t+\ddt)  =&\int \!\!\text{d}Y_t \, \wp_\text{ost}(Y_t)\,\hat{M}_\text{I}(Y_t)\rho(t)\hat{M}^\dagger_\text{I}(Y_t)\nonumber\\
	 = &\, \rho(t)+\mathcal{D}[\hat{c}]\rho(t)\ddt+\tfrac{1}{4}\hat{c}^\dagger\hat{c}\rho(t)\hat{c}^\dagger\hat{c}\ddt^2.
\end{align}
We can see that the resulting average dynamics agrees with the Lindblad evolution Eq.~\eqref{stateupdate}, to only first order in $\ddt$, while the ${\cal O}(\ddt^2)$ term is just a part of the double Lindblad operator Eq.~\eqref{LindHigh}.

Even though the average dynamics Eq.~\eqref{eq-aveito1} is linear in $\rho(t)$ to second order in $\ddt$, the trace-preserving condition (C3) of the It\^o measurement operator is only satisfied to the first order in $\ddt$. This can be verified via the completeness condition in Eq.~\eqref{CompletenessRelation}.  Since we have the Kraus-form operator, we can directly compute the completeness condition as,
\begin{align}\label{CPIto}
  \int\!\!\dd Y_t \, \wp_\text{ost}(Y_t)\, \hat{M}^\dagger_\text{I}(Y_t)\hat{M}_\text{I}(Y_t) = &\, \hat{1}+\tfrac{1}{4}\big(\hat{c}^\dagger\hat{c}\big)^2\ddt^2,
	\end{align}
which turns out to have a term of the order ${\cal O}(\ddt^2)$ in addition to the identity. In other words, from Eq.~\eqref{CPIto}, we show that the It\^o measurement operator $\hat M_{\rm I}(Y_t)$ is only trace preserving (C3) to first order in $\ddt$. \blk %Basing CPness on the map is clearer, as it generalizes to the ``non-kraus'' approach. %The reader is reminded that, to generalise the measurement operator, one can include the Hamiltonian evolution by adding $-i\hat{H}\ddt$ to the map which correspond to the first order in $\ddt$ of the unitary evolution expansion, $-i[\hat{H},\bullet]$.

\subsubsection{It\^o approach Method II: Nonlinear}

The above violation of trace-preservation of the linear It\^o approach --- which is needed for purposes such as quantum state smoothing --- is not typically an issue with quantum state filtering. That is because this can be treated using a nonlinear, exactly trace-preserving, method. Here the standard implementation of this It\^o approach is to approximate the readout's PDF as Gaussian, where the readout can be written as a sum of its informative and noisy parts,
\begin{eqnarray}\label{Mrecord}
Y_t \Delta t &=& \mu_\text{I} \Delta t +  \Delta W_{\rm I}.
\end{eqnarray}
Here $\Delta W_{\rm I}$ is the well-known zero-mean Wiener increment and $\mu_{\rm I}$ is the mean readout. In order to find a consistent $\mu_\text{I}$ and verify the moments of $\Delta W_{\rm I}$, one can use Eq.~\eqref{actualprobIto} and Eq.~\eqref{Mrecord} to show that
\begin{align}
\mu_\text{I} = &\, \langle \hat{c}+\hat{c}^\dagger\rangle + {\cal O}(\ddt),\\
\sigma_\text{I}^2 =&\, 1 / \ddt + {\cal O}(\ddt^0),
\end{align}
following the definitions of $\mu$ and $\sigma$ in Eqs.~\eqref{eq-mu} and \eqref{eq-sigma}, using the notation $\langle \bullet \rangle \equiv \text{Tr}[\bullet\rho(t)]$ and keeping terms to lowest order in $\ddt$. We can also verify that
\begin{subequations}\label{Noise}
\begin{eqnarray}
\Eavg{\Delta W_{\rm I}} &=&0,\\ \label{Noiseb}
\Eavg{\Delta W_{\rm I}^2} &=& \ddt, \\
\Eavg{\Delta W_{\rm I}^4} &=& 3\ddt^2,
\end{eqnarray}
\end{subequations}
describing statistics of the  Wiener process. Therefore, we can approximate the readout's PDF as a Gaussian distribution with the mean $\mu_\text{I}$ and variance $\sigma_{\rm I}$, i.e.,
\begin{align}\label{eq-probyito2}
    \wp_{\rm I, g}(Y_t|\rho(t)) = \left(\frac{\ddt}{2\pi}\right)^{1/2}\!\! \exp\left[-\left(Y_t - \langle \hat c + \hat c^\dagger\rangle\right)^2\ddt/2\right].
\end{align}
 %where the subscript `g' indicates that the readout PDF. 
It is interesting to note that this guessed PDF is different from the PDF from Kraus operators, $\wp_{\rm I, k}(Y_t|\rho(t))$ in Eq.~\eqref{actualprobIto}.
%The It\^o interpretation is commonly applied to stochastic processes which involve Gaussian white noise and the Wiener increment.
%If one were to write a probability distribution of $Y_t$ given its form in Eq.~\eqref{Mrecord}, one gets
%\begin{align}
%\wp(Y_t) = \left(\frac{\Delta t}{2\pi}\right)^{1/2} \exp\left[(Y_t-\mu_t)^2\Delta t/2\right]
%\end{align}
 
From Table~\ref{Summary-Error}, we only need to verify the conditions (B) and (C2). Using the Gaussian-approximated PDF of $Y_t$ and the statistical moments of $\Delta W_{\rm I}$ above, we can show that the average (unconditioned) quantum trajectory, following the second line of Eq.~\eqref{eq-Lindblad}, gives a different result from the previous case to second order in $\ddt$. By expanding the conditioned state, $\tilde{\cal J}[\hat{M}_\text{I}(Y_t)]\rho(t)$ in Eq.~\eqref{eq-Lindblad}, keeping terms of the order $\Delta t^2$ and $\Delta W_{\rm I}^4$, writing $Y_t$ in terms of $\Delta W_{\rm I}$, and adopting the moments of $\Delta W_{\rm I}$ in Eqs.~\eqref{Noise}, we get: 
\begin{align}\label{Ito-nonlinear}
\rho(t+\ddt)&= \Eavg{ \tilde{\cal J}[\hat{M}_\text{I}(Y_t)]\rho(t)}\nonumber\\
&=\rho(t)+\mathcal{D}[\hat{c}]\rho(t)\,\ddt+\Big[\tfrac{1}{4}\hat{c}^\dagger\hat{c}\rho(t)\hat{c}^\dagger\hat{c}  \nonumber\\
&+\hat{c}\rho(t)\hat{c}^\dagger \big( \langle \hat{c}+\hat{c}^\dagger \rangle ^2 - 2\langle \hat{c}^\dagger\hat{c} \rangle \big)\nonumber \\
& + \rho(t)\big(  2\langle \hat{c}^\dagger \hat{c} \rangle ^2 + 3 \langle \hat{c}^\dagger\hat{c} \rangle \langle \hat{c}+\hat{c}^\dagger \rangle^2 -\tfrac{1}{2}\langle \hat{c}^\dagger \hat{c}^2\nonumber\\
& + (\hat{c}^\dagger)^2\hat{c} \rangle \langle \hat{c}+\hat{c}^\dagger \rangle -\tfrac{1}{4} \langle (\hat{c}^\dagger\hat{c})^2\rangle + \langle \hat{c}+\hat{c}^\dagger\rangle^4 \big)\nonumber\\
&- [\rho(t)\hat{c}^\dagger + \hat{c}\rho(t)] \big( -\langle \hat{c}+\hat{c}^\dagger \rangle^3 + 2 \langle \hat{c}+\hat{c}^\dagger \rangle \langle \hat{c}^\dagger \hat{c} \rangle  \nonumber \\
& +\tfrac{1}{2}\langle \hat{c}^\dagger \hat{c}^2 + (\hat{c}^\dagger)^2\hat{c} \rangle \big)\Big]\ddt^2 + \mathcal{O}(\ddt^3).
\end{align}
Once again, it is seen that this unconditioned evolution only aligns with the Lindblad evolution (B) to the first order in $\ddt$. For the convex-linearity (C2), we can infer from Eq.~\eqref{Ito-nonlinear} that its right-hand side is nonlinear in $\rho$ for terms of order $\ddt^2$. The nonlinear terms arise from the trace of the dominator, which is $\text{Tr}[\mathcal{J}[\op M_\text{I}(Y_t)]\rho(t)]\ne 1$ even if $\text{Tr}[\rho(t)]=1$. Thus, the condition (C2) for the It\^o map is satisfied only to the first order in $\ddt$.

\subsection{Rouchon-Ralph approach}\label{RRApproach}
%In RR paper, with their earlier paper, they are proposing the new way of filtering that corresponding to SME with positive and trace preserving, solving problems in SME integration with finite time step, stability and submartingale...comment: SME comparison, 

Let us move on to the next existing approach, an extended version of the It\^o approach. Proposed by Rouchon and Ralph~\cite{Rouchon2015} as a more efficient quantum filtering for continuous weak measurement, it is based on the Euler-Milstein stochastic simulation method~\cite{Milstein1995}. The proposed technique makes the stochastic increment strongly convergent to first order in $\ddt$ as opposed to the conventional It\^o approach, which is weakly convergent to $\ddt$~\cite{Rouchon2015}. %The technique also ensures the Hermiticity and positivity of the quantum state and can be implemented for systems with feedback controls and inefficient measurements.  
Following the notation used in the previous subsection for the single Lindblad operator $\hat{c}$ and assuming that the technique can be used with the coarse-grained measurement record $Y_t$, we  write the Rouchon-Ralph measurement operator as
\begin{align}\label{MRR}
\hat{M}_\text{R}(Y_t) =\, \,&\hat{1}-\tfrac{1}{2}\hat{c}^\dagger\hat{c}\ddt+\hat{c}Y_t\ddt -\tfrac{1}{2}\hat{c}^2\big(\ddt-Y_t^2\ddt^2\big),
\end{align}
where the last term is an addition to the It\^o measurement operator, Eq.~\eqref{MIto}. This term corresponds to the correction following the Euler-Milstein method and will vanish if one naively applies the It\^o rule, i.e., $Y_t^2\ddt^2 \approx \ddt$, making $\hat{M}_\text{R}\approx\hat{M}_\text{I}$. It is also interesting to note that, what we call, the Rouchon-Ralph (RR) operator, Eq.~\eqref{MRR}, can be alternatively derived from the interaction unitary operator as described in Section~\ref{sec-constructU}, using Eqs.~\eqref{eq-U1}~and~\eqref{eq:measopgen}, where all terms up to the order ${\cal O}(\dt^{3/2})$ are included (also replacing $\dt$ and $y_s$ with $\ddt$ and $Y_t$, respectively). In the following subsubsections, similar to the analyses in the It\^o case, we present the two ways of writing the readout's PDF used in computing the average trajectory and its properties, and show how this RR map satisfy the hierarchy conditions. %The additional terms in the Rouchon-Ralph version of the measurement operator are exactly from $\tfrac{1}{2}\bra{Y_t}\hat{c}^2(\dB^\dagger)^2\ket{0} =\tfrac{1}{2}\hat{c}^2(Y_t^2\ddt^2-\ddt)$. Therefore, applying the It\^o rule, this terms will vanish.

\subsubsection{RR approach Method I: Linear}
Given the unnormalized measurement operator $\hat M_{\rm R}$ in Eq.~\eqref{MRR} and the ostensible probability in Eq.~\eqref{Xost} as in the It\^o case, the Kraus operator for the Rouchon-Ralph (RR) approach is written as
\begin{align}
    \hat K_{\rm R}(Y_t) = \sqrt{\wp_{\rm ost}(Y_t)} \hat M_{\rm R}(Y_t),
\end{align}
which, following the definition of the readout's PDF in Eq.~\eqref{eq-probygen}, leads to
\be \label{RouchonDist}
\wp_{\rm R, k}(Y_t | \rho(t)) = \wp_{\rm ost}(Y_t) {\rm Tr}\left\{ {\cal J}[\hat{M}_{\rm R}(Y_t)]\rho(t) \right\}.
\ee
As before, we can compute the average unconditioned dynamics to second order in $\ddt$ to obtain
\begin{align}\label{RRmethod1}
	\rho(t+\ddt) =&\int\!\!\text{d}Y_t \, \wp_\text{ost}(Y_t)\, \hat{M}_\text{R}(Y_t)\rho(t)\hat{M}^\dagger_\text{R}(Y_t) \nonumber \\
	=&\,\rho(t)+\mathcal{D}[\hat{c}]\rho(t)\ddt \\
	&+\big[\tfrac{1}{4}\hat{c}^\dagger\hat{c}\rho(t)\hat{c}^\dagger\hat{c}+\tfrac{1}{2}\hat{c}^2\rho(t)(\hat{c}^\dagger)^2\big]\ddt^2 + {\cal O}(\ddt^3),\nonumber
 \end{align}
which only agrees with the Lindblad evolution (B) in Eq.~\eqref{stateupdate} only to first order in $\ddt$. For the trace-preserving condition (C3), we compute the completeness relation in Eq.~\eqref{CompletenessRelation}, and find that
\begin{align}\label{eq-averou1}
 \int\!\!\text{d}Y_t \, \wp_\text{ost}(Y_t)\, &\hat{M}^\dagger_\text{R}(Y_t)\hat{M}_\text{R}(Y_t)   \\
 &  =\hat{1}+\big[\tfrac{1}{4}(\hat{c}^\dagger\hat{c})^2+\tfrac{1}{2}(\hat{c}^\dagger)^2\hat{c}^2\big]\ddt^2+{\cal O}(\ddt^3),\nonumber
\end{align}
which shows that the RR measurement operator satisfies (C3) only to first order in $\ddt$, same as for the It\^o case. Interestingly, even though there is an extra positive term of ${\cal O}(\ddt^2)$ in the completeness condition, beyond the standard It\^o case in Eq.~\eqref{CPIto}, we will see later in Section~\ref{MapCompare} that this approach results in a slight improvement in estimating individual trajectories compared to the original It\^o approach.

%The Rouchon-Ralph measurement operator is also generalized to include the unitary dynamics following Ref.~\cite{Rouchon2015}, by adding $-i\hat{H}\ddt -\tfrac{1}{2}\hat{H}^2\ddt^2$. %We note that, by taking the time-continuum limit, $\ddt \rightarrow \dt$, one still get back the It\^o SME as in Eq.~\eqref{ItoSME}. 

\subsubsection{RR approach Method II: Nonlinear}

%Similar to the It\^o case, the exact readout's PDF Eq.~\eqref{RouchonDist} can be inconvenient if used in practice. 
In the work by Rouchon and Ralph~\cite{Rouchon2015} and the one from Rouchon works~\cite{ROUCHON2022}, they preferred using the Gaussian approximation for the measurement readout's PDF, where the measurement readout was written as
\begin{align}
Y_t \ddt = \mu_{\rm R} \ddt + \Delta W_{\rm R},
\end{align}
where the statistical properties of $\Delta W_{\rm R}$ are exactly those of $\Delta W_{\rm I}$ in Eqs.~\eqref{Noise}.
%\begin{subequations}\label{NoiseRR}
%\begin{eqnarray}
%\Eavg{\Delta W_{\rm R}} &=&0,\\ \label{Noiseb}
%\Eavg{\Delta W_{\rm R}^2} &=& \ddt \\
%\Eavg{\Delta W_{\rm R}^4} &=& 3\ddt^2,
%\end{eqnarray}
%\end{subequations}
%resulting in $\mu_{\rm R} = \mu_{\rm I}$ and $\sigma_{\rm R}^2 = \sigma_{\rm I}^2= 1/\ddt$~. 
Using these statistics, we compute the average evolution from Eq.~\eqref{eq-Lindblad} and get
\begin{align}\label{RRmethod2}
\rho(t+\ddt)&=\Eavg{ \tilde{\cal J}[\hat{M}_\text{R}(Y_t)]\rho(t)},\\
&=\Eavg{ \tilde{\cal J}[\hat{M}_\text{I}(Y_t)]\rho(t)} + \Big[ \tfrac{1}{2}\hat{c}^2\rho(t)(\hat{c}^\dagger)^2\nonumber\\
& + \rho(t)\Big(-\tfrac{1}{2}\langle \hat{c}^2(\hat{c}^\dagger)^2\rangle -\tfrac{1}{2} \langle \hat{c}^2+(\hat{c}^\dagger)^2\rangle \langle \hat{c}^\dagger+ \hat{c}\rangle^2\nonumber \\
&- \langle (\hat{c}^\dagger)^2\hat{c} + \hat{c}^\dagger\hat{c}^2\rangle \langle \hat{c}+\hat{c}^\dagger\rangle +\tfrac{1}{2}\langle \hat{c}^2+(\hat{c}^\dagger)^2\rangle^2 \Big) \nonumber\\
&+ [\rho(t)\hat{c}^\dagger + \hat{c}\rho(t)]\Big( -\langle \hat{c}^2+(\hat{c}^\dagger)^2 \rangle \langle \hat{c}+\hat{c}^\dagger \rangle   \nonumber \\
&-\langle \hat{c}^\dagger \hat{c}^2 + (\hat{c}^\dagger)^2\hat{c} \rangle \Big) -\hat{c}\rho(t)\hat{c}^\dagger\langle \hat{c}^2+(\hat{c}^\dagger)^2 \rangle
\nonumber\\
& + [\hat{c}^2\rho(t) +\rho(t)(\hat{c}^\dagger)^2] \Big( \tfrac{1}{2}\langle \hat{c}+\hat{c}^\dagger \rangle ^2 \nonumber\\
&- \tfrac{1}{2}\langle \hat{c}^2 +(\hat{c}^\dagger)^2 \rangle -\langle \hat{c}^\dagger\hat{c}\rangle \Big)\Big]\, \ddt^2 + \mathcal{O}(\ddt^3), \nonumber
\end{align}
where we can see that, again, the result only agrees with Lindblad evolution (B) to first order in $\ddt$, same as in the It\^o case. We note that the error terms of order ${\cal O}(\ddt^2)$ from \hyperlink{method1}{Method I}~Eq.~\eqref{RRmethod1} and \hyperlink{method2}{Method II}~Eq.~\eqref{RRmethod2} are different from each other. Moreover, one can see that the non convex-linear in $\rho(t)$ terms appear in ${\cal O}(\ddt^2)$, failing to satisfy the condition (C2) at $\mathcal{O}(\ddt^2)$.
 
 %%%%%%%%%%%%%%
\subsection{Guevara-Wiseman approach}\label{GWApp}

% in Ivonne's paper, it is mainly discussing the new map that can satisfy the completely positive condition, better than the previous maps. However, the new terms are kinda of heuristically obtained.

The last existing approach we consider is that introduced by Guevara and Wiseman~\cite{Guevara2020}. Since the It\^o map satisfies the complete positivity only to first order in $\ddt$, they considered adding extra terms to the measurement operator such that it would satisfy the completeness relation to the order of $\ddt^2$. Their so-called complete positivity map for quantum trajectories was proposed for both jumps and diffusive measurement records and was implemented in the quantum state smoothing~\cite{Guevara2015,Chantasri_2019}. Using our notation for consistency, the Guevara-Wiseman measurement operator for a single Lindblad channel $\op c$ and the coarse-grained record is given by
\begin{eqnarray}\label{GWMO}
\hat{M}_\text{G}(Y_t)&=&\hat{1}+\big(Y_t\hat{c}-\tfrac{1}{2}\hat{c}^\dagger\hat{c}\big)\ddt-\tfrac{1}{8}(\hat{c}^\dagger\hat{c})^2\ddt^2.
\end{eqnarray}
We note that this operator has one extra term,  $-\frac{1}{8}(\hat{c}^\dagger\hat{c})^2\ddt^2$, in addition to the It\^o operator. This extra term was proposed as a correction term, in order to remove the non-zero term in the second order in $\ddt$ in the It\^o completeness condition Eq.~\eqref{CPIto}. As before, we compare the properties of the operator computed from two methods.

\subsubsection{GW approach Method I: Linear}

We follow the original derivation in Ref.~\cite{Guevara2020} and use the operator $\hat M_{\rm G}$ in Eq.~\eqref{GWMO} and its ostensible probability in Eq.~\eqref{Xost} to construct the Kraus operator for the Guevara-Wiseman (GW) approach 
\begin{align}\label{GWMO2}
    \hat K_{\rm G}(Y_t) = \sqrt{\wp_{\rm ost}(Y_t)} \hat M_{\rm G}(Y_t)
\end{align}
which leads to the readout's PDF as
\begin{align}\label{actualprobGue}
\wp_{\rm G, k}(Y_t|\rho(t)) =\wp_{\rm ost}(Y_t){\rm Tr}\left\{{\cal J}[\op{M}_{\rm G}(Y_t)] \rho(t)\right\}.
\end{align}
Using the above to compute the average dynamics to second order in $\ddt$, we obtain
\begin{align}\label{eq-aveGueWis}
	\rho(t+\ddt) = &\int\!\!\text{d}Y_t\wp_\text{ost}(Y_t)\hat{M}_\text{G}(Y_t)\rho(t)\hat{M}^\dagger_\text{G}(Y_t)\\
	=&\, \rho(t)+\mathcal{D}[\hat{c}]\rho(t)\ddt+\tfrac{1}{4}\mathcal{D}[\hat{c}^\dagger\hat{c}]\rho(t)\ddt^2 + {\cal O}(\ddt^3).\nonumber
\end{align}
We can see that, the average dynamics still only match with the exact Lindblad evolution (B), in Eq.~\eqref{stateupdate} to the first order in $\ddt$. However, for the completeness condition, we have
\begin{align}
\int\!\!\text{d}Y_t\,\wp_\text{ost}(Y_t)\,\hat{M}^\dagger_\text{G}(Y_t)\hat{M}_\text{G}(Y_t) =&\, \hat{1}+\mathcal{O}(\ddt^3),
\end{align}
which confirms the purpose of GW measurement operator that it should be trace preserving (C3) to second order in $\ddt$, better than in the previous two approaches. 

\subsubsection{GW approach Method II: Nonlinear}

In the original proposal Ref.~\cite{Guevara2020}, the authors also suggested a consistent way to approximate the readout's PDF as a Gaussian distribution that has the same statistics as the PDF in Eq.~\eqref{actualprobGue} to second order in $\ddt$. In order to achieve that, the mean and the variance of the Gaussian distribution have to be modified.
Similarly to before, let us assume 
\begin{align}
Y_t\ddt =\mu_\text{G}\ddt + \Delta W_\text{G},
\end{align}
we can now compute the modified mean and variance using Eq.~\eqref{actualprobGue} to get
\begin{align}
\mu_\text{G} & = \langle\hat{c}+\hat{c}^\dagger \rangle -\tfrac{1}{2} \langle \hat{c}^\dagger\hat{c}^2+\hat{c}(\hat{c}^\dagger)^2 \rangle \ddt +\mathcal{O}(\ddt^2),\nonumber\\
\sigma^2_\text{G} &= \tfrac{1}{\ddt} + \big[ 2\langle\hat{c}^\dagger\hat{c} \rangle - \langle\hat{c}^\dagger+\hat{c} \rangle^2 \big]  +\mathcal{O}(\ddt),
\end{align}
noting that there were typos in the original paper Ref.~\cite{Guevara2020} which should be replaced by our new calculations. We can also compute the new statistics of $\Delta W_{\rm G}$ from Eq.~\eqref{actualprobGue}, which gives
\begin{align}
\Eavg{\Delta W_{\rm G}} &=\Eavg{\Delta W_{\rm G}^3} = 0,\nonumber\\
\Eavg{\Delta W_{\rm G}^2} &= \ddt + \big( 2\langle\hat{c}^\dagger\hat{c} \rangle - \langle \hat{c}^\dagger+\hat{c} \rangle^2 \big) \ddt^2 +\mathcal{O}(\ddt^3),\nonumber\\
\Eavg{\Delta W_{\rm G}^4} &= 3\ddt^2 +\mathcal{O}(\ddt^3).
\end{align}
These moments are consistent with a Gaussian distribution, with a higher-order correction to the variance compared to the standard It\^o $\Delta W_\text{I}$. We can use the above properties to compute the average dynamics, i.e.,
\begin{align}
\rho(t+\ddt)& = \Eavg{\tilde{\cal J}[\hat{M}_\text{G}(Y_t)]\rho(t)},\\
&=\rho(t)+\mathcal{D}[\hat{c}]\rho(t)\ddt +\tfrac{1}{4}\mathcal{D}[\hat{c}^\dagger\hat{c}]\rho(t)\ddt^2 + \mathcal{O}(\ddt^3),\nonumber
\end{align}
which exactly coincides with Eq.~\eqref{eq-aveGueWis} using the PDF from the Kraus operator. Despite failing to agree with high-order Lindblad evolution, this measurement operator ensures the convex-linearity condition (C2) to $\mathcal{O}(\ddt^2)$. Since (C1) and (C2) are automatically satisfied (see Table.~\ref{Summary-Error}), this means that the GW-map $\op M_\text{G}$ generates valid average quantum evolution correct to $\mathcal{O}(\ddt^2)$.  %We would like to note that this consistency can be achieved even for the It\^o's or Rouchon-Ralph's maps, by modifying the mean and variance of the readout's Gaussian distribution, similar to what was shown above.

%To generalize the measurement operator to include the unitary dynamics, instead of adding the expansion of the unitary operator, Guevara and Wiseman consider the measurement and unitary dynamics separately~\cite{Guevara2020}. They prefer adding the unitary operation, $\mathcal{U}_{\ddt}[\bullet]=e^{-i\hat{H}\ddt}\bullet e^{+i\hat{H}\ddt}$, on top of the measurement operation. This agrees with the operation, $-i[\hat{H},\bullet]$, for the first order expansion in $\ddt$. 

%%%%%%%%%%%%%%%%%%%%%%
% CONSTRUCTING MEASUREMENT OPERATOR FROM PHYSICAL
%%%%%%%%%%%%%%%%%%%%%%%
\section{Constructing High-order Completely Positive Map}\label{Construct}
\par We have shown in the previous section that none of the existing approaches, even with two different ways of computing the readout's average, gives a perfect agreement with the Lindblad evolution (B) and two of them fail to generate valid average quantum evolution (C), to orders higher than $\ddt$. In this section, we propose a measurement operator that satisfy all conditions (B) and (C1)-(C3) to second order in $\ddt$. We derive such operator based on series expansion of the exact Kraus operator in Eq.~\eqref{eq-extkraus}. %considers the primary system coupled to its environment. We leave the system of interest as a general quantum system described by a density matrix $\rho$ coupled to a bosonic environment initialized in vacuum state via a Lindblad operator $\hat c$. 
%We have shown that different orders (or terms) of expansions of the interacting operator give two different measurement operators, including the It\^o map and the Rouchon-Ralph map. 
Similar to other approaches, we only focus on the case with a single Lindblad channel and $\op H=0$ for simplicity.

Let us consider the exact Kraus operator and its expansion in Eqs.~\eqref{eq-extkraus} and \eqref{eq-U1}, but now keep terms from its expansion to ${\cal O} [|\dB_t|^4]$. %We therefore tailor expanded terms by selecting only ones with the desired orders and coefficients, such that the additional contribution of ${\cal O}(\ddt^2)$ in the average dynamics agree perfectly with the expansion of the Lindblad equation to this order in Eq.~\eqref{stateupdate0}, as well as the conditions C1-C3 (see Appendix~\ref{app-highorder} for the detailed derivation). With the selected terms, the Kraus operator in Eq.~\eqref{eq-U1} becomes
%\begin{align}\label{U2expans}
%\hat{K}(y_s) =& \, \op K_1(y_s) + \ \bra{y_t} \tfrac{1}{8}(\op{c}^\dagger\op{c})^2(\dB_t\dB_t^\dagger)^2|0\rangle \nonumber\\
%& - \langle y_t|\tfrac{1}{8}\op{c}^\dagger\op{c}^2\dB_t(\dB_t^\dagger)^2 +\tfrac{1}{4}\op{c}\,\op{c}^\dagger\op{c}\,(\dB_t^\dagger\dB_t\dB_t^\dagger ) |0 \ra,\nonumber\\
%& + {\cal O}[|\dB_t|^m],
%\end{align}
%where \red $\op K_1(y_s)$ \blk  represents the terms explicitly shown in Eq.~\eqref{eq-U1}, which are of the order: $\dB_t^2$, $\dB_t\dB_t^\dagger$, or $(\dB_t^\dagger)^2$, and below. 
As before, using the annihilation and creation properties of the bath operator $\dB_t$ and the wavefunction in Eq.~\eqref{eq-Hermite}, we obtain a high-order measurement operator denoted as $\op M_2(y_s)$ for an infinitesimal measurement results $y_s$, where its full form is presented in Appendix~\ref{app-highorder}. We then follow an approach similar to that of Eq.~\eqref{MIto}, deriving a map for a finite $\ddt$ by discretizing time into \(m=\ddt/\dt\) segments, multiplying $\op M_2(y_s)$ $m$ times, and taking limit \(m \rightarrow \infty\) (see Appendix~\ref{app-highorder} for more detail). This yields our proposed coarse-grained operator $\op M_\text{W}(Y_t)$ for a finite $\ddt$ as
\begin{multline}\label{Mwwc1}
\hat{M}_{\rm W}(Y_t)=\hat{1}-\tfrac{1}{2}(\hat{c}^2+\hat{c}^\dagger\hat{c})\ddt+\tfrac{1}{8}\big(\hat{c}^\dagger\hat{c}\big)^2\ddt^2\\+\big[\hat{c}\ddt-\tfrac{1}{4}\big(\hat{c}^\dagger\hat{c}^2+\hat{c}\hat{c}^\dagger\hat{c}\big)\ddt^2\big]Y_t+\tfrac{1}{2}\hat{c}^2\ddt^2Y_t^2.
\end{multline}
This is one of our main results in this work. We see that our operator can be written in terms of the RR operator, $\op M_\text{R}(Y_t)$ in Eq.~\eqref{MRR}, with three additional (high-order) terms: $\hat{M}_\text{W}(Y_t)=\hat{M}_\text{R}(Y_t)+\tfrac{1}{8}\big(\hat{c}^\dagger\hat{c}\big)^2\ddt^2-\tfrac{1}{4}\big(\hat{c}^\dagger\hat{c}^2+\hat{c}\hat{c}^\dagger\hat{c}\big)\ddt^2Y_t$. As we show in the following, we can use this high-order Kraus operator with two types of the readout's PDF to give the same average dynamics, and satisfying the conditions (B) and (C) to second order in $\ddt$.

\subsubsection{Higher-order approach Method I: Linear}

Following the similar manner as in the existing approaches, using the operator in Eq.~\eqref{Mwwc1} and the ostensible probability in Eq.~\eqref{Xost}, we can write the Kraus operator for our high-order approach as 
\begin{align}
    \hat K_{\rm W}(Y_t) = \sqrt{\wp_{\rm ost}(Y_t)} \hat M_{\rm W}(Y_t),
\end{align}
which leads to the readout's PDF,
\begin{align}\label{eq-probW}
\wp_{\rm W, k}(Y_t | \rho(t)) = \wp_{\rm ost}(Y_t){\rm Tr}\left\{ {\cal J}[\hat{M}_{\rm W}(Y_t)] \rho(t) \right\}.
\end{align}
Using this PDF, we compute the average dynamics to the second order in $\ddt$ and find that
\begin{multline}\label{eq-aveWWW}
\rho(t+\ddt)=\int\!\!\text{d}Y_t \, \wp_\text{ost}(Y_t)\hat{M}_{\rm W}(Y_t) \rho(t)\hat{M}^\dagger_{\rm W}(Y_t)  \\
	=\rho(t)+\mathcal{D}[\hat{c}]\rho(t)\ddt+\tfrac{1}{2}\mathcal{D}^2[\hat{c}]\rho(t)\ddt^2 +\mathcal{O}(\ddt^3),
\end{multline}
the average dynamics now are convex-linear and match exactly with the Lindblad solution (B) in Eq.~\eqref{stateupdate} to the second order in $\ddt$. From the implication in Fig.~\ref{fig:diagram}(b), it follows that the condition (C) should be satisfied to the same order. Indeed we can show that 
\begin{align}\label{CPwwc}
	\int \!\! \text{d}Y_t \, \wp_\text{ost}(Y_t)\hat{M}^\dagger_{\rm W}(Y_t)\hat{M}_{\rm W}(Y_t)  = & \hat{1}+\mathcal{O}(\ddt^3),
\end{align}
or the trace preservation (C3) is also exactly satisfied to second order in $\ddt$. We will reproduce this with the other (nonlinear) method in the next following.

\subsubsection{Higher-order approach Method II: Nonlinear}

As in the Guevara-Wiseman approach, we can approximate the readout's PDF as a Gaussian distribution, but still obtain the same readout's statistics as the PDF in Eq.~\eqref{eq-probW} to the second order in $\ddt$. Similarly to other approaches, the readout is of the form
\begin{align}
Y_t \ddt =\mu_\text{W} \ddt + \Delta W_\text{W},
\end{align}
where the modified average and variance of the readout are computed from Eq.~\eqref{eq-probW}, which give  
\begin{multline}
\mu_\text{W}  = \langle\hat{c}+\hat{c}^\dagger \rangle + \tfrac{1}{4} \langle  (\hat{c}^\dagger)^2\hat{c} + \hat{c}^\dagger\hat{c}^2 - \hat{c}\hat{c}^\dagger\hat{c} - \hat{c}^\dagger\hat{c}\hat{c}^\dagger \rangle \ddt  +\mathcal{O}(\ddt^2),\\
\sigma^2_\text{W}
 = \tfrac{1}{\ddt} + \big[ 2\langle\hat{c}^\dagger\hat{c} \rangle - \langle\hat{c}^\dagger+\hat{c} \rangle^2 + \langle \hat{c}^2 +(\hat{c}^\dagger)^2 \rangle \big]  +\mathcal{O}(\ddt).
\end{multline}
We also find that the innovation $\Delta W_{\rm W}$ has the consistent ensemble-average properties:
\begin{align}
\Eavg{\Delta W_{\rm W}} &=\Eavg{\Delta W_{\rm W}^3} = 0,\nonumber\\
\Eavg{\Delta W_{\rm W}^2} &= \ddt + \big[ 2\langle\hat{c}^\dagger\hat{c} \rangle - \langle\hat{c}^\dagger+\hat{c} \rangle^2 + \langle \hat{c}^2 +(\hat{c}^\dagger)^2 \rangle  \big] \ddt^2 ,\nonumber\\
\Eavg{\Delta W_{\rm W}^4} &= 3\ddt^2.
\end{align}
Therefore, using the above properties, we can derive the average dynamics from
\begin{multline}
\rho(t+\ddt)=\Eavg{\tilde{\cal J}[\hat{M}_\text{W}(Y_t)]\rho(t)},\\
=\rho(t)+\mathcal{D}[\hat{c}]\rho(t)\ddt+\tfrac{1}{2}\mathcal{D}^2[\hat{c}]\rho(t)\ddt^2+ \mathcal{O}(\ddt^3),
\end{multline}
which is exactly equal to Eq.~\eqref{eq-aveWWW}. We have shown that, again, the proposed high-order operator Eq.~\eqref{Mwwc1} results in the average evolution in perfect agreement with the Lindblad solution (B) and satisfy the conditions (C1)-(C3)  to second order in $\ddt$.

%%%%%%%%%%%%%%
% Qubit example
%%%%%%%%%%%%%%
 \section{Qubit Example with exact Maps}\label{qubitExample} 

The proposed map has demonstrated high accuracy in generating average evolutions. The next natural question that arises is how accurate it is at the level of the conditioned (trajectory) evolution. In order to address the question, one needs an exact quantum trajectory as a benchmark. Here, we consider two examples of qubit measurement for which the conditioned trajectories can be solved exactly (or to high order of accuracy). These are the qubit $z$-measurement (Hermitian measurement, $\hat{c}\propto \hat{\sigma}_z$) and the measurement of qubit's fluorescence (non-Hermitian measurement, $\hat{c}\propto \hat{\sigma}_-$). Both have already been tested on various architectures of superconducting qubit experiments ~\cite{Vijay2012,Murch2013,Weber2014,chantasri2016,Shay2016noncom,Jordan2016,Philippe2020}. In the following subsections, we derive corresponding ``exact'' Kraus operators for the two examples.  We will use these operators as a benchmark for exact individual trajectories in Section~\ref{MapCompare}.

%\ach{put this paragraph somewhere} Considering a two-level system as the quantum system of interest coupled to a bosonic bath via a dissipative channel described by a Lindblad operator $\hat c = -i \sqrt{\gamma} \hat\sigma_-$, which is related to the lowering operator and $\gamma$ is the dissipative rate.

%The latter can be obtained the exact map by solving the It\^o differential equations which, however, there is no explicit form of the measurement operator~Ref[]. Nevertheless, the two exact measurement operators do not have a general form for arbitrary Lindblad operators as the existing maps do. 

 %%%%%%%%%%%%%%
\subsection{Measurements with Hermitian Lindblad operators (e.g., $z$-measurement)}\label{ExactZmap}
We begin by considering a weak continuous measurement of a Hermitian observable. Measurements using Hermitian Lindblad operators align with the traditional von Neumann concept of measurement with quantum observables, where post-measurement states collapse to eigenstates of the observables over time. These measurements can range from weak (imprecise) to strong (projective). The measurement operator for this case can be motivated by Bayesian inference and statistical principles. This technique was originally proposed for solid-state qubits, utilizing quantum dots and quantum point contacts as measurement devices~\cite{Korotkov1999,Korotkov2001,Korotkov2002,Ruskov2003}. It has since been adopted in continuous quantum measurement applications in superconducting qubit experiments. In this subsection, we present the derivation for any observable $\hat A = \sum_j a_j \ket{a_j}\bra{a_j}$, where $\ket{a_j}$ and $a_j$ are eigenstates and their corresponding eigenvalues of the observable. 
 
Since we consider the weak continuous measurement, an important assumption is the central limit theorem, where the statistics of imprecision can be described by Gaussian distribution. Therefore, the probability distribution of a measurement readout $Y_t$, given that the state was in the eigenstate $\ket{a_j}$, is
\begin{align}\label{eq-proby-bay}
\wp(Y_t | a_j)=\left(\frac{\gamma \ddt}{\pi}\right)^{1/2}\exp\left[-\frac{\gamma}{2}\left( \tfrac{Y_t}{\sqrt{2\gamma}}  -a_j\right)^2 \ddt\right],
\end{align}
where the measurement record is scaled such that the PDF is a Gaussian function around the mean value $a_j$.  %Noting that one can find the probability distribution of the usual record as
%\begin{align}\label{eq-proby-bay}
%\wp(X | a_j)=\left(\frac{\gamma \ddt}{\pi}\right)^{1/2}\exp\left[-\frac{\gamma}{2}( \tfrac{X}{\sqrt{2\gamma}}  -a_j)^2 \ddt\right],
%\end{align}
%where the relationship between the scaled record to the usual record is given by $r_t=\sqrt{\tfrac{1}{2\gamma}}X$. 
We note that the coefficient $\gamma$ will be later related to a coupling rate. Following the original idea in Ref.~\cite{Korotkov1999}, one can write the Kraus operator for this Hermitian measurement as~\cite{Jacobs2006}
\begin{eqnarray}\label{Bayesian}
\hat{K}_{\rm H}(Y_t)&=&\sum_j \sqrt{\wp(Y_t|a_j)}\ket{a_j}\bra{a_j},
\end{eqnarray}
which is a matrix with diagonal elements given by the square roots of the readout probability distribution. Using Eq.~\eqref{eq-probygen}, the operator Eq.~\eqref{Bayesian} trivially gives the probability of the measurement result,
\begin{align}
\wp(Y_t | a_j) = {\rm Tr}[\hat K_{\rm H}(Y_t) \ket{a_j}\bra{a_j} \hat K_{\rm H}^\dagger(Y_t)],
\end{align}
which is the exact PDF as in Eq.~\eqref{eq-proby-bay}. 

\par We can show that this measurement operator gives the average dynamics agreeing with the Lindblad solution to infinite orders in $\ddt$, by showing that 
\begin{eqnarray}\label{LinBayZ}
	\rho(t+\ddt) &=& \Eavg{{\tilde{\cal J}[\hat K_{\rm H}(Y_t)]\rho(t)}} \nonumber \\&=& \int\text{d}Y_t\hat{K}_\text{H}(Y_t)\rho(t)\hat{K}^\dagger_\text{H}(Y_t)\nonumber\\
	&=& e^{\ddt\mathcal D[\sqrt{\gamma/2}\hat A]\bullet}\rho(t),
	\end{eqnarray}
where the third line comes from the fact that $\hat K_{\rm H}$ and $\hat A$ are both diagonal in the eigenstate basis $\{ |a_j\rangle \}$. The coupling $\gamma$ is defined in the Lindblad operator $\op c=\sqrt{\gamma/2}\op A$. Eq.~\eqref{LinBayZ} justifies that this set of Kraus operators reproduces the Lindblad evolution to infinite order of $\ddt$. An example of measurements with Hermitian Lindblad operators is the qubit measurement in $z$-basis, where $\hat{A} = \hat\sigma_z$, with two eigenstates: $|a_{+1}\ra = |\rm e\ra$ and $|a_{-1}\ra = |\rm g\ra$, the excited and ground states of the qubit, respectively.

%Noting that although the Hermitian Kraus operator Eq.~\eqref{Bayesian} satisfies both the complete positivity and unravelling condition to infinite orders in $\ddt$, this is only exact when $\hat H = 0$. For the case when $\hat H \ne 0$, the state update is calculated operationally via the superoperator $\mathcal{U}_{\ddt}$. 

%We can conclude without proof that, when the three superoperators do not commute, i.e., $[\hat H , \hat A] \ne 0$, the completeness relation and the unconditioned state update should no longer be satisfied to infinite order of $\ddt$. \blk

%Since we consider imperfect measurements, we can think of the losing information as averaging all possible lost readouts, which is
%\begin{eqnarray}
%\mathcal{O}_\gamma&=&\int\text{d}r_u\hat{M}_{r_u}\rho\hat{M}^\dagger_{r_u},
%\end{eqnarray}{}
%where we call it a dephasing term and operate this on the off-diagonal elements of $\rho$.
%The state update evolution of qubit is given by
%\begin{eqnarray}\label{bayeupdate}
%\rho(t+\ddt)&=&\mathcal{O}_\gamma\mathcal{U}_{\delta t}\mathcal{M}_{r_n}[\rho(t)],
%\end{eqnarray}{}
%where  $\mathcal{U}_{\delta t}[\rho(t)]=e^{-i\hat{H}\delta t}\rho(t) e^{i\hat{H}\delta t}$ is a unitary operation and $\mathcal{M}_{r_n}[\rho(t)]=\frac{\hat{M}_{r_n}\rho(t)\hat{M}_{r_n}^\dagger}{\Tr(\hat{M}_{r_n}\rho(t)\hat{M}_{r_n}^\dagger)}$.

  %%%%%%%%%%%%%%
 \subsection{Measurement of qubit fluorescence}\label{sec-qflu}
\par The second example is qubit fluorescence measurement, which examines the energy relaxation or the transition of states of a qubit, i.e., from its excited state $\ket{\text{e}}$ to ground state $\ket{\text{g}}$. This results in an emission of a photon at the energy difference between the two levels. The photon emitted from the system can be detected by a measurement device, which, in this work, we assume a homodyne measurement. This type of measurement corresponds to the Lindblad operator $\hat c \propto \hat\sigma_-$, which is a qubit's lowering operator, i.e., non-Hermitian measurement. For non-Hermitian Lindblad operators, there is no general exact form of measurement operators as in Eq.~\eqref{Bayesian}, but we can derive one for a qubit case. In the following, we start with deriving the exact Kraus operator, which is in terms of a weighted readout integral. Then, for it to be useful in analytical calculation later, we need to make an approximation to obtain a ``nearly exact'' operator. %\red Other relevant approximated versions, such as the one in Ref.~\cite{Philippe2020}, will be detailed in Appendix~\ref{flu-bayesian}.\blk

\subsubsection{Qubit fluorescence measurement: Exact operator}

\par The measurement operator for qubit fluorescence measurement for a finite time $\ddt$ can be derived analytically by integrating the differential equation, which describes how the operator changes in an infinitesimal time. Consider a time $s+\dd s$, where $s$ is introduced as a dummy variable for time and $\dd s$ is an infinitesimal time. An operator at $s+\dd s$ can be decomposed as $\hat{M}(s+\dd s)=\hat{M}(\dd s)\hat{M}(s)$. This leads to a differential equation for the operator,
\begin{align}\label{diff-mat-flu}
\dd \hat{M}&= \hat M(s+\dd s) - \hat M(s) = [\hat{M}(\dd s)-\hat{1}]\hat{M}(s),
\end{align}
Since $\dd s$ is infinitesimal, we can replace $\hat M(\dd s)$ by the It\^o operator in Eq.~\eqref{MIto} as it should be exact up to first order in $\dd s$.
We then use $\hat{c}=\sqrt{\gamma}\hat{\sigma}_-$ for the fluorescence measurement and show that the differential equation Eq.~\eqref{diff-mat-flu} becomes
\begin{align}\label{eq-fluelement}
\begin{pmatrix}\dd m_{00}&\dd m_{01}\\ \dd m_{10} &\dd m_{11}\end{pmatrix}&=\begin{pmatrix} -\tfrac{1}{2}\gamma \dd s &0\\ \sqrt{\gamma}y_s\dd s &0\end{pmatrix} \begin{pmatrix} m_{00}(s)& m_{01}(s)\\ m_{10}(s) & m_{11}(s)\end{pmatrix},
\end{align}
where we have used $m_{jk}(s)$ for $j,k \in \{0,1\}$ as the four elements of the matrix $\hat{M}(s)$. 
%\begin{subequations}
%\begin{align}
%\dd m_{00}&=-\tfrac{1}{2}\gamma m_{00}(s)\dd s,\\
%\dd m_{01}&=-\tfrac{1}{2}\gamma m_{01}(s)\dd s,\\
%\dd m_{10}&=\sqrt{\gamma} m_{00}(s)\dd s,\\
%\dd m_{00}&=\sqrt{\gamma} m_{01}(s)\dd s.
%\end{align}
%\end{subequations}

To obtain the operator for a finite-time $\ddt$, each element in Eq.~\eqref{eq-fluelement} should be integrated from a time of interest $t$ to $t+\ddt$, with an initial condition $\hat{M}(t = 0)=\hat{1}$. Therefore, one obtains the exact (unnormalized) measurement operator for the fluorescence case as 
\begin{align}\label{FluExact}
\hat{M}_\text{F,ex}(X_t)&=\begin{pmatrix}e^{-\gamma\ddt/2}&0\\ \sqrt{\gamma}X_t &1\end{pmatrix},
\end{align}
where
\begin{align}\label{defX}
    X_t&\equiv \int_t^{t+\ddt} e^{-\gamma s/2}y_s\dd s
\end{align}
%Note that the bottom left element of Eq.~\eqref{FluExact} involves 
is a weighted readout integral from time $t$ to $t+\ddt$. To construct the Kraus operator for this case, we need to find the ostensible probability $\wp_\text{ost}(X_t)$ such that 
\begin{align}\label{exactKrausFlu}
    \op K_\text{F,ex}(X_t)&=\sqrt{\wp_\text{ost}(X_t)}\op M_\text{F,ex}(X_t). 
\end{align}
The probability can be obtained via a multi-variable integration of the ostensible probability, $\wp(y_s)$, for infinitesimal measurement results as defined in Eq.~\eqref{ostprob-dt} (see Appendix~\ref{moments-ZY} for the detailed derivation). First, we find that $X_t$ has zero mean, i.e., $\text{E}\{ X_t\} = 0$, which is expected as $X_t$ is a linear combination of zero-mean Gaussian variables, $y_s$. Second, we find that the variance is given by $\sigma_{X_t}^2=\text{E}\{ X_t^2\}- \text{E}\{ X_t\}^2 = 2e^{-\gamma(t+\ddt/2)}\sinh(\gamma\ddt/2)/\gamma$. Thus, we can construct the ostensible probability for $X_t$ as
\begin{align}\label{Xost-exact}
    \wp_\text{ost}(X_t)&=\frac{1}{\sqrt{2\pi\sigma_{X_t}^2}}\exp[-X_t^2/(2\sigma_{X_t}^2)].
\end{align} 

The exact Kraus operator with $\wp_\text{ost}(X_t)$ above is not trivial for analytical calculations. This brings us to approximate the weighted readout integral $X_t$. In the previous work~Refs.~\cite{Jordan2016,Philippe2020}, this integrated readout was approximated to $\sqrt{\gamma}X_t \approx \sqrt{\gamma}Y_t\ddt$, which is a Taylor expansion of the exponential weight $e^{-\gamma s/2}$ to only the zeroth order in $s$ (we show in Appendix~\ref{flu-bayesian} that the resulting map only satisfies the condition (B) to first order in $\ddt$, which is lower than what we want). Therefore, in order to benchmark exact individual trajectories, we need a more accurate map, considering higher-order terms in $s$. This will be derived in the following.

\subsubsection{Qubit fluorescence measurement: Nearly exact operator}

To calculate errors in the quantum trajectory, it is sufficient to consider an approximated version that can still capture the statistics of measurement results to high orders in $\ddt$. Let us first expand the exponential function in $X_t$ in the integral of Eq.~\eqref{defX} to first order in $s$, i.e., $e^{-\gamma s/2}= 1-\gamma s/2$. One can rearrange terms, giving 
\begin{align}\label{eqinty}
X_t&= \int_t^{t+\ddt} \!\! e^{-\gamma s/2} y_s  \dd s\nonumber\\
&\approx \int_t^{t+\ddt} \!\! \big[ 1 -\tfrac{\gamma}{2} (t+\tfrac{\ddt}{2})-\tfrac{\gamma}{2}\left(s-t-\tfrac{\ddt}{2}\right) \big]y_s \dd s\nonumber\\ 
&= [1-\tfrac{\gamma}{2}(t+\tfrac{\ddt}{2})]Y_t \ddt  -\tfrac{\gamma}{2}Z_t,
\end{align}
where $Y_t$ is as defined in Eq.~\eqref{coarse-grain-weight}, and
\begin{align}\label{YZdis}
Z_t &\equiv \int_t^{t+\ddt} \!\! [s-(t+\tfrac{\ddt}{2})] y_s\dd s,
\end{align}
which is another integrated-record variable. We will show later in Appendix~\ref{moments-ZY} that, by defining the new variable this way, $Y_t$ and $Z_t$ are statistically independent.
%where it retains the leading order of expansion in \( s \). By expressing it this way, we will demonstrate later that it holds significant statistical utility. 
The measurement operator thus reads
\begin{align}\label{approx-map-flu}
\hat{M}_{\rm F}(Y_t, Z_t)&=\begin{pmatrix}e^{-\gamma\ddt/2}&0\\ \sqrt{\gamma}[1-\tfrac{\gamma}{2}(t+\tfrac{\ddt}{2})]Y_t \ddt -\tfrac{\gamma^{3/2}}{2}Z_t&1\end{pmatrix},
\end{align}
which will be used as a benchmark for generating quantum trajectories. 

Since there are now two record variables, $Y_t$ and $Z_t$, we need to, again, find their statistical properties and moments of their ostensible probabilities. %These can be obtained via a multi-variable integration of the ostensible probability, $\wp(y_s)$, as defined in Eq.~\eqref{ostprob-dt}. 
With the first and second moments, we can evaluate PDF of the variable $Z_t$ (see Appendix~\ref{moments-ZY} for the detailed derivation of multi-variable integration). First, we find that $Z_t$ has zero mean, i.e., $\text{E}\{ Z_t\} = 0$.  Next, the variance is given by $\text{E}\{ Z_t^2\}- \text{E}\{ Z_t\}^2 = \Delta t^3/12$. We can further show that the co-variance of $Y_t$ and $ Z_t$ is zero, meaning that they are statistically independent, i.e., $\text{E}\{ Y_t Z_t \} = 0$. We therefore construct an ostensible probability for $Z_t$ as
\begin{align}\label{Zost}
\wp_\text{ost}(Z_t)&=\sqrt{\tfrac{6}{\pi \ddt^3}}\exp(-6Z_t^2/\ddt^3).
\end{align}
Notice that the variable $Z_t$ has the dimension of $\ddt^{3/2}$. The nearly-exact Kraus operator for the qubit's fluorescence measurement is thus
\begin{align}\label{eqflukraus}
    \hat K_{\rm F}(Y_t,Z_t) = \sqrt{\wp_{\rm ost}(Y_t)\wp_{\rm ost}(Z_t)} \hat M_{\rm F}(Y_t,Z_t),
\end{align}
using the ostensible probabilities defined in Eqs.~\eqref{Xost} and \eqref{Zost}. Note that these only depend on the time step $\ddt$, so the ensemble of the records is easier to generate than using $\wp_\text{ost}(X_t)$. \blk  We will use this form of the Kraus operator for the analytical calculation in the next section.
 
 \begin{table*}[t!]
 \setlength{\tabcolsep}{8pt} % Default value: 6pt
 \renewcommand{\arraystretch}{1.8} % Default value: 1
 \begin{center}
\begin{tabular}{|c | c | c|}
	\hline
	Methods & Average trace distance compared to $\hat{K}_{\text{H}}(Y_t)$& Average trace distance compared to $\hat{K}_{\text{F}}(Y_t, Z_t)$ \\  
	\hline 
	$\hat K_{\text{I}}$ &$\frac{7\sqrt{\pi}}{48}(\gamma\ddt)^{3/2}+\mathcal{O}(\ddt^{5/2}) \approx 0.2585(\gamma\ddt)^{3/2}$ & $\frac{1}{2\sqrt{6\pi}}(\gamma\ddt)^{3/2}+\mathcal{O}(\ddt^{5/2}) \approx 0.1152(\gamma\ddt)^{3/2}$ \\ [1ex] 
	\hline
	$\hat K_{\text{R}}$ &$\frac{(1+e^3)\sqrt{\pi}}{12e^3}(\gamma\ddt)^{3/2}+\mathcal{O}(\ddt^{5/2})\approx 0.1551(\gamma\ddt)^{3/2} $& $\frac{1}{2\sqrt{6\pi}}(\gamma\ddt)^{3/2}+\mathcal{O}(\ddt^{5/2}) \approx 0.1152(\gamma\ddt)^{3/2}$ \\ [1ex] 
	\hline
	$\hat K_{\text{G}}$ &$ \frac{7\sqrt{\pi}}{48}(\gamma\ddt)^{3/2}+\mathcal{O}(\ddt^{5/2}) \approx 0.2585(\gamma\ddt)^{3/2}$& $\frac{1}{2\sqrt{6\pi}}(\gamma\ddt)^{3/2}+\mathcal{O}(\ddt^{5/2}) \approx 0.1152(\gamma\ddt)^{3/2}$ \\ [1ex] 
	\hline
	$\hat K_{\text{W}}$ &$ \ \frac{(4+e^{3/2})\sqrt{\pi}}{48e^{3/2}}(\gamma\ddt)^{3/2}+\mathcal{O}(\ddt^{5/2}) \approx 0.0699(\gamma\ddt)^{3/2}$&  $\frac{1}{4\sqrt{6\pi}}(\gamma\ddt)^{3/2}+\mathcal{O}(\ddt^{5/2}) \approx 0.0576(\gamma\ddt)^{3/2}$\\ [1ex] 
	\hline
\end{tabular}
\end{center}
\caption{Analytical results of the average trace distance integrated over all possible states and measurement results defined in Eq.~\eqref{DOB} for the two examples: the qubit $z$-measurement (left column) and the qubit fluorescence measurement of qubit fluorescence (right column). %Here, the subscripts I, R, G and W refer to $\hat{K}_{\rm I}, \hat{K}_{\rm R}, \hat{K}_{\rm G}$ and $\hat{K}_{\rm W}$, respectively.
}
\label{TDS}
\end{table*} 

%%%%%%%%%%%%%%%%%%%%%%
%Map Comparison
%%%%%%%%%%%%%%%%%%%%%%%
\section{Individual Quantum Trajectory Comparison}\label{MapCompare}

%In Sections~\ref{ExisApps} and~\ref{Construct}, we derived the average dynamics and their properties according to the three conditions for both existing measurement operators and our proposed high-order operator. 
In this section, we use the two qubit examples from the previous section—the qubit \( z \)-measurement and the qubit fluorescence measurement—to investigate the valid quantum trajectory condition (A) for all measurement operators. We do this by examining how closely the individual trajectories they generate match the exact trajectories produced by the exact (and nearly exact) operators. To measure the closeness between any two quantum states, we use the trace distance and compute an average over all possible states and measurement records. %that is independent of specific measurement records and the quantum system's state. 

Consider how an arbitrary quantum state $\tilde{\rho}$ will change after one finite-time step $\Delta t$, where the finite-time (coarsed-grained) measurement result $Y_t$ is obtained. Denoting an exact (or nearly exact) Kraus operator by $\hat K_{\rm ex}(\tilde Y_t)$ for relevant record variables $\tilde Y_t$, which can be either $Y_t$ for the qubit $z$-measurement Eq.~\eqref{Bayesian} or $\{Y_t, Z_t\}$ for the qubit fluorescence in Eq.~\eqref{eqflukraus}, we compute the benchmark conditioned quantum state as
\begin{align}\label{exactMap}
\rho_{\rm ex}(\tilde Y_t, \tilde \rho)= \frac{\hat{K}_{\rm ex}(\tilde Y_t)\, \tilde \rho\, \hat{K}_{\rm ex}^\dagger(\tilde Y_t)}{\text{Tr}[\hat{K}_{\rm ex}(\tilde Y_t)\,\tilde \rho\, \hat{K}_{\rm ex}^\dagger(\tilde Y_t)]}.
\end{align}
If one instead uses an operator of the four $\hat K_{\rm A}$, where ${\rm A} \in \{ {\rm I, R, G, W} \}$ defined in Eqs.~\eqref{MIto},\eqref{MRR},\eqref{GWMO} and \eqref{Mwwc1}, respectively, only the coarse-grained $Y_t$ is measured and the conditioned state becomes
\begin{align}\label{GenMap}
\rho_{\rm A}(Y_t, \tilde \rho)= \frac{\hat{K}_{\rm A}(Y_t)\, \tilde \rho\, \hat{K}_{\rm A}^\dagger(Y_t)}{\text{Tr}[\hat{K}_{\rm A}(Y_t)\,\tilde \rho\, \hat{K}_{\rm A}^\dagger(Y_t)]},
\end{align}
as different estimates of the exact state. 
Therefore, we define the average trace distance for an approach A,
\begin{eqnarray}\label{DOB}
D_{\rm A} &=& \tfrac{1}{2}\int \!\! \dd \mu_{\rm H}(\tilde \rho) \!\! \int \!\! \dd \tilde Y_t\, \wp_{\rm ex}(\tilde Y_t|\tilde \rho) 
 \,\, \text{Tr}\big| \rho_{\rm A}- \rho_{\rm ex}\big|.
\end{eqnarray}
Note that the trace distance between the state $\rho_{\rm A} = \rho_{\rm A}(Y_t,\tilde\rho)$ in Eq.~\eqref{GenMap} and the exact state $\rho_{\rm ex} = \rho_{\rm ex}(\tilde Y_t,\tilde\rho)$ in Eq.~\eqref{exactMap} is averaged  over the measurement results $\tilde Y_t$, with its probability weight computed from the exact measurement operator, $\wp_{\rm ex}(\tilde Y_t|\tilde\rho) = {\rm Tr}[\hat K_{\rm ex}(\tilde Y_t)\tilde\rho \hat K_{\rm ex}^\dagger(\tilde Y_t)]$.
The trace distance is then also averaged over possible states $\tilde{\rho}$ to make the average distance independent of any initial state. For convenience, we consider $\tilde \rho = \ket{\psi}\bra{\psi}$ as a pure state of a single qubit, i.e., $|\psi\rangle =\cos{(\theta/2)}\ket{0}+e^{i\phi}\sin{(\theta/2)}\ket{1}$, and the integral is over the Harr-measure, $\dd \mu_{\rm H}=\dd \phi\sin(\theta)\dd \theta/(4\pi)$.

\subsection{Comparison for qubit $z$-measurement}
We begin with the qubit \( z \)-measurement. In this case, the Lindblad operator is \(\hat{c} = \sqrt{\gamma/2} \hat{\sigma}_z\), and the exact measurement operator is given by \(\hat{K}_\text{H}\) in Eq.~\eqref{Bayesian}. Here, there is only one record variable to compute the exact map, \(\tilde{Y}_t = Y_t\). Therefore, we calculate the average over the coarse-grained measurement results with their exact probability distribution \(\wp_\text{ex}(Y_t |\tilde{\rho}) = \text{Tr} [\hat{K}_\text{H}(Y_t) \tilde{\rho} \hat{K}_\text{H}^\dagger(Y_t)]\). Using the average trace distance in Eq.~\eqref{DOB} and the exact PDF, we perform analytical calculations using \emph{Mathematica} and obtain results for different measurement operators shown in Table~\ref{TDS} (left column). %For the integration over the Haar measure, we transform it to an integral over the qubit's Bloch sphere with $\theta \in[0, \pi]$ and $\phi \in [0,2\pi]$. 
To simplify the analytical calculation, we had to Taylor-expand the integrand of Eq.~\eqref{DOB} to the $4$th order in $\gamma$. We chose $\gamma$ instead of $\ddt$ as our expansion parameter to consistently track the expansion orders of the time-related variables.

From the results in Table~\ref{TDS}, we see that all the average trace distances are of order of $\ddt^{3/2}$, even that for the higher-order map $\op K_\text{W}$ we introduced. However, they have different pre-factors and our proposed operator $\op K_\text{W}$ gives the smallest distance to the exact trajectories. The operators from It\^o and Guevara-Wiseman approaches give states with the largest average distance from the exact state.

\subsection{Comparison for qubit's fluorescence measurement}

For the second example, the Lindblad operator is $\hat{c}=\sqrt{\gamma}\hat{\sigma}_-$, with $\gamma$ representing the qubit's fluorescence (decay) rate. We use the nearly exact version of the measurement operator derived in Eqs.~\eqref{approx-map-flu} and~\eqref{eqflukraus}, where we need the two types of coarse-grained measurement results $Y_t$ and $Z_t$. Therefore, the average over the measurement results in Eq.~\eqref{DOB} now becomes the average over both $Y_t$ and $Z_t$, where the probability weight is replaced by
\begin{align}
\wp_{\rm ex}(\tilde Y_t | \tilde \rho)= {\rm Tr}[\hat K_{\rm F}(Y_t,Z_t) \, \tilde\rho\, \hat K^\dagger_{\rm F}(Y_t,Z_t)],
\end{align}
given $\tilde Y_t = \{Y_t, Z_t\}$ and using the Kraus operator defined in Eq.~\eqref{eqflukraus}. The statistical independence of $Y_t$ and $Z_t$ allows us to integrate the two variables directly with $\wp_{\rm ex}(\tilde Y_t | \tilde \rho)$.  We performed the analytical calculations similarly to the $z$-measurement case and the results are presented in Table~\ref{TDS} (right column).

Again, our proposed measurement operator $\hat{K}_\text{W}$ outperforms other approaches, this time by a pre-factor $1/2$. We also note that $\hat{K}_\text{I}$ and $\hat{K}_\text{R}$ give exactly the same result for this example because the two maps coincide when $\hat{c}^2=\hat{\sigma}_-^2=0$.

%------------------------------------------
% DISCUSSION AND CONCLUSION
%------------------------------------------
\section{Conclusion}\label{conclusion}
As measurement records from real experiments are usually obtained
with finite time resolution ($\ddt$), existing tools used in processing the records to
get quantum trajectories can easily lead to numerical errors in resulting quantum states. In this work, we have presented a systematic approach to evaluate the accuracy of quantum trajectory and to analyze
what should be a good (Kraus) map based on the hierarchy of conditions: (A) valid quantum trajectory, (B) Lindblad evolution, and (C) valid average quantum evolution. The last can be broken down into three conditions: (C1) complete positivity, (C2) convex-linearity, and (C3) trace preservation. We considered the accuracy of these conditions at  the second order in $\ddt$, i.e., one order higher than traditional treatments. We then reviewed existing maps from the literature, finding that all considered approaches failed to satisfy the Lindblad evolution (B) and two of them, $\hat K_{\rm I}(Y_t)$ and $\hat K_{\rm R}(Y_t)$, failed to generate the valid average quantum
evolution (C), to second order in $\ddt$.

We therefore introduced a technique to construct measurement operators from a
unitary system-bath interaction in the interaction-frame 
expanded to fourth order
in the bath operator. For a single Lindblad operator and  \(\hat{H}=0\), we have constructed a higher-order map $\hat K_{\rm W}(Y_t)$ that satisfies all the desired conditions (B) and (C), to second order in \(\ddt \). We also showed that the higher-order map outperforms other approaches, giving the smallest distance to the exact trajectories, which is the strongest condition (A), at least for the two qubit examples: the qubit \(z\)-measurement and the qubit fluorescence measurement. Analytical results are summarized in Table~\ref{TDS}.

\begin{table}[t!]
\setlength{\tabcolsep}{5pt} % Default value: 6pt
 \renewcommand{\arraystretch}{1.5} % Default value: 1
\begin{tabular}{|l|c|c|c|c|}
	\hline 
	Conditions &  $\hat{K}_{\rm I}$ & $\hat{K}_{\rm R}$ & $\hat{K}_{\rm G}$ &  $\hat{K}_{\rm W}$  \\ 
     \hline
	(A) VQT (two examples) & \multicolumn{4}{c|}{$\mathcal{O}
(\ddt)$}  \\  
	\hline
 (B) Lindblad solution & \multicolumn{3}{c|}{$\mathcal{O}
(\ddt)$} & $\mathcal{O}
(\ddt^2)$  \\ 
     \hline
	(C) VAQE (Method I \& II)  & \multicolumn{2}{c|}{$\mathcal{O}
(\ddt)$} &\multicolumn{2}{c|}{$\mathcal{O}
(\ddt^2)$} \\  
	\hline
\end{tabular}
\caption{\label{Summary-all} Summary of the order of accuracy in $\ddt$ for the three hierarchy conditions (A), (B), and (C) for the existing maps. Note that VQT stands for the valid quantum trajectory and VAQE stands for valid average quantum evolution.
}
\end{table}

We can also summarize the accuracy of the hierarchy conditions (A), (B), and (C), which can be computed analytically for all approaches we considered in Table~\ref{Summary-all}. From bottom to top rows, the condition (C) is satisfied to ${\cal O}(\ddt^2)$ if one uses the Guevara-Wiseman map, $\hat K_{\rm G}(Y_t)$, or our proposed map, $\hat K_{\rm W}(Y_t)$, while the condition (B) is satisfied only for $\hat K_{\rm W}(Y_t)$. For the strongest condition (based on the two qubit examples, see Table~\ref{TDS}), all approaches give the average trace distance to the exact quantum trajectories at the order of \(\mathcal{O}(\ddt^{3/2})\) and above, i.e., the accuracy is only to \(\mathcal{O}(\ddt)\). As mentioned in the previous paragraph that our proposed map outperforms all others with the smallest prefactors of the average trace distance. Notably, the systematic hierarchy of assessing the map is consistent with the one-way implication diagram.

Given our analysis, we believe that our proposed map $\hat K_{\rm W}(Y_t)$ will be useful when a high-accuracy calculation of quantum evolution is needed and a time resolution of measurement records cannot be assumed infinitesimal. Our immediate future work is to implement the new map to measurement records from the experiment of fluorescence measurement of transmon qubits, Ref~\cite{NagFor2016}, where the calculation of high-accuracy quantum trajectories is needed in multi-parameter estimation using sequential Monte Carlo method~\cite{RalMas2017,Chattamas2024}. Moreover, we will also explore the generalization to multiple Linblad channels and non-zero system's Hamiltonian in the forthcoming companion paper~\cite{WWC2023}. A final aspect that can be explored is the possibility of including the integrated record $Z_t$, Eq.~\eqref{YZdis}, in a Kraus map to enhance the accuracy of individual trajectory calculation.

%------------------------------------------
% Acknowledgement
%------------------------------------------
\section{Acknowledgement}
NW thanks S. Suwanna for support during the time he worked on this project at Mahidol University, Thailand. NW was also supported by the Development and Promotion of Science and Technology Talents Project Thailand (DPST). AC acknowledges the support of the Griffith University Postdoctoral Fellowship scheme and Australian Research Council Centre of Excellence Program CE170100012. This research has received funding support from the NSRF via the Program Management Unit for Human Resources and Institutional Development, Research and Innovation (Thailand) [grant number B39G670018].
%------------------------------------------
% Appendix
%------------------------------------------
\appendix 

%%%%%%%%%%%%%%
%%%% Full page %%%%%
%%%%%%%%%%%%%%

\section{Complete positivity for averaged SMEs}\label{CPofSME}

In this section, we show the violation of the complete positivity condition (C1) of the averaged SME. We start with an example of a two-qubit system initialized in a maximally entangled state, \(\rho_\text{AB}(t) = \frac{1}{2} (\ket{00} + \ket{11})(\bra{00} + \bra{11})\), where the first space encodes a qubit subsystem A, and the second space encodes an ancillary state (subsystem B).

The first-order averaged SME can be expressed for the combined system as
\begin{align}
    \rho_\text{AB}(t+\ddt)&=\rho_\text{AB}(t)+\ddt\mathcal{D}[\op c \otimes \op 1_\text{B}],
\end{align}
where $\op 1_\text{B}$ is the identity operator in the ancillary state space and $\op c$ acts on the qubit subsystem A. One example that the complete positivity condition (C1) is violated is the non-Hermitian measurement of qubit fluorescence. The corresponding Lindblad operator is defined $\op c =\sqrt{\gamma/2}\op\sigma_-$, where $\op\sigma_- = \frac{1}{2}(\op\sigma_x-i\op\sigma_y)$ acting as a lowering operator on the qubit. Substituting the superoperator $\mathcal{D}$ to the joint state, we have
\begin{align}\label{CP-updatestate}
\rho_\text{AB}(t+\ddt)=& \frac{1}{2}\begin{pmatrix}1&0&0&1-\gamma\ddt/2\\ 0&\gamma\ddt&0&0\\ 0&0&0&0\\ 1-\gamma\ddt/2&0&0&1-\gamma\ddt \end{pmatrix}.
\end{align}
The lowest eigenvalue of the matrix in Eq.~\eqref{CP-updatestate} is
\begin{align}
    \lambda&=-\frac{1}{4}[\gamma\ddt+\sqrt{2(2-\gamma\ddt(2-\gamma\ddt))}-2]\approx -\frac{(\gamma\ddt)^2}{4}.\nonumber
\end{align}
The condition (C1) is clearly violated as the lowest eigenvalue is negative, with the size of $\ddt^2$.

\section{Derivation of the It\^o operator}\label{app-itomap}

We can construct the It\^o measurement operator for a finite time increment $\ddt$ from the measurement operator of an infinitesimal record $\hat M(y_s)$ found in Eqs.~\eqref{eq-U1} and \eqref{eq:measopgen},
\begin{align}\label{M1-ys}
\hat M_1(y_s) = \hat{1}-\tfrac{1}{2}\hat{c}^\dagger\hat{c}\dt+\hat{c}y_s\dt +\tfrac{1}{2} \hat{c}^2 (y_s^2\dt^2-\dt).
\end{align}
by combining the operators for all $m = \ddt /\dt$ infinitesimal records defined as $\{ y_s : s \in \{t, t+\dt, ..., t+(m-1)\dt\}$, keeping terms up to only ${\cal O}(\dt^2)$. We first compute the product of the unnormalized operators and take the continuum limit to get
\begin{multline}
\hat M_\text{I}(Y_t) = \lim_{m\rightarrow \infty} \hat M_1(y_{t+(m-1)\dt})  \cdots \hat M_1(y_{t+\dt}) \hat M_1(y_{t}) \\
= \lim_{m\rightarrow \infty} \!\!\! \prod_{s = t}^{t+(m-1)\dt}\left[\hat 1 - \tfrac{1}{2} \hat c^\dagger \hat c \dt + \hat c y_s \dt +\tfrac{1}{2} \hat{c}^2 (y_s^2\dt^2-\dt) \right], \\
= \, \hat 1 -\tfrac{1}{2} (\hat c^\dagger \hat c + \hat c^2 )\Delta t + \lim_{m\rightarrow \infty} \!\!\! \sum_{s=t}^{t+(m-1)\dt} \!\!\! (\hat c y_s \dt + \tfrac{1}{2} \hat c^2 y_s^2 \dt^2 ).
\end{multline}
In the last line above, in the mean square limit, we can show that the infinite sum, ${\cal Y} \equiv \sum_s y_s^2 \dt^2$, has a stochastic variance converging to zero, 
\begin{align}
   \lim_{m\rightarrow \infty} {\rm E}\{  {\cal Y}^2 \} -& {\rm E} \{ {\cal Y} \}^2  \nonumber \\
   =& \!\! \lim_{m\rightarrow \infty} \dt^4 \sum_s {\rm E} \left\{ y_s^4 \right\} - \dt^4\sum_s {\rm E}\left\{ y_s^2 \right\}^2, \nonumber \\
   =& \!\! \lim_{m\rightarrow \infty} 4 \mu^2 \frac{\ddt^3}{m^2} + 2 \frac{\ddt^2}{m} = 0,
\end{align}
where we have used ${\rm E}\{ y_s \} = \mu$ and ${\rm E}\{y_s^2 \} = \mu^2 + 1/\dt$. This means that we can replace the infinite sum ${\cal Y}$ with its stochastic mean, 
\begin{align}\label{sumys2}
    \lim_{m\rightarrow \infty} {\rm E}\{  {\cal Y} \}  = \lim_{m\rightarrow \infty} \mu^2 \frac{\ddt^2}{m} + \ddt = \ddt,
\end{align}
and we obtain the unnormalized measurement operator for a finite time $\ddt$ using the It\^o approach as
\begin{align}
    \hat M_\text{I}(Y_t) = \hat 1 + \tfrac{1}{2} \hat c^\dagger \hat c \ddt +\hat c Y_t \ddt,
\end{align}
where we have defined a coarse-grained record
\begin{align}\label{coarse-grain-weight-1}
Y_t \equiv \frac{1}{\ddt}\! \int_t^{t+\ddt} \!\!{\rm d}s\, y_s 
 = \lim_{m \rightarrow \infty} \frac{1}{\ddt} \!\!\!  \sum_{s=t}^{t+(m-1)\dt} \!\! y_s \dt.
\end{align}
We can also find a normalized factor for this operator by convolution, from a product of $\wp_{\rm ost}(y_s)$ for $y_s$ satisfying Eq.~\eqref{coarse-grain-weight}. Because they are all Gaussian functions, we find that a normalized factor is $\sqrt{\wp_{\rm ost}(Y_t)}$, with a new ostensible probability 
\begin{align}
\wp_{\rm ost}(Y_t) &= \left(\frac{\ddt}{2\pi}\right)^{1/2}\!\! \exp(-Y_t^2\ddt/2).
\end{align} 
\smallskip

\section{Derivation of the high-order measurement operator} \label{app-highorder}

In this section we will show the detailed derivation of the high-order completely positive map from Eq.~\eqref{Mwwc1} which are the high-order expansion of the unitary $\hat{U}_{t+\dt, t}$. Let us consider the coupling unitary operator in Eq.~\eqref{eq-uopint} and define 
\begin{eqnarray}
\op{U}_{t+\dt,t}&=&\exp(\op \beta)=\sum_{k=0}^\infty \frac{\hat{\beta}^k}{k!},
\end{eqnarray}
where $\op{\beta}\equiv\op{c}\dB^\dagger-\op{c}^\dagger\dB$. 
The Taylor-expanded $\hat{\beta}$ up to $4^{\rm th}$ order (or equivalently to $\mathcal{O}[|\dB|^4]$) and compute matrix elements of $\bra{y_s}\op U_{t+\dt, t}\ket{0}$ as
\lipsum[0]
\begin{widetext}
\begin{eqnarray}\label{u-expansion}
\bra{y_s} \op{\beta}^0 |0 \ra&=& \sqrt{\wp_{\rm ost}(y_s)}\\% new line
\bra{y_s} \op{\beta}^1 |0 \ra&=&\bra{y_s} \op{c}\dB^\dagger |0 \ra =\sqrt{\wp_{\rm ost}(y_s)} y_s\op{c}\dt \nonumber\\% new line
\bra{y_s} \op{\beta}^2 |0 \ra &=&\bra{y_s} \op{c}^2(\dB^\dagger)^2-\op{c}^\dagger\op{c}\dB\dB^\dagger |0 \ra = \sqrt{\wp_{\rm ost}(y_s)}\big[ \op{c}^2(y^2_s\dt-1)-\op{c}^\dagger\op{c}\big]\dt \nonumber\\% new line
\bra{y_s} \op{\beta}^3 |0 \ra &=&\bra{y_s} \op{c}^3(\dB^\dagger)^3-\op{c}^\dagger\op{c}^2\dB(\dB^\dagger)^2 -\op{c}\op{c}^\dagger\op{c}\dB^\dagger\dB\dB^\dagger |0 \ra = \sqrt{\wp_{\rm ost}(y_s)}\big[ \op{c}^3(y^3_s\dt-3y_s)-2y_s\op{c}^\dagger\op{c}^2-y_s\op{c}\op{c}^\dagger\op{c}\big]\dt^2\nonumber\\% new line
\bra{y_s} \op{\beta}^4 |0 \ra &=&\bra{y_s} \op{c}^4(\dB^\dagger)^4-\op{c}^\dagger\op{c}^3\dB(\dB^\dagger)^3 -\op{c}\op{c}^\dagger\op{c}^2\dB^\dagger\dB(\dB^\dagger)^2-\op{c}^2\op{c}^\dagger\op{c}(\dB^\dagger)^2\dB\dB^\dagger\nonumber\\
&+&(\op{c}^\dagger\op{c})^2(\dB\dB^\dagger)^2+(\op{c}^\dagger)^2\op{c}^2\dB^2(\dB^\dagger)^2 |0 \ra \nonumber \\
&=& \sqrt{\wp_{\rm ost}(y_s)} \big[ \op{c}^4(y^4_s\dt^2-6y^2_s\dt+3)-(3\op{c}^\dagger\op{c}^3+2\op{c}\op{c}^\dagger\op{c}^2 + \op{c}^2\op{c}^\dagger\op{c})(y_s^2\dt-1)+(\op{c}^\dagger\op{c})^2+2(\op{c}^\dagger)^2\op{c}^2\big]\dt^2,\nonumber 
\end{eqnarray}
where we have omitted the terms consisting the rightmost operator $\dB_t$ as $\dB_t\ket{0}=0$.

We obtain the unnormalized high-order measurement operator with infinitesimal records as
\begin{align}\label{wwch}
    \op M_2(y_s)&=\op M_1(y_s)+\left[\frac{1}{24}(\op{c}^\dagger\op{c})^2+\frac{1}{12}(\op{c}^\dagger)^2\op{c}^2 + \op M_\text{h}(y_s)\right]\dt^2,
\end{align}
where $\op M_1(y_s)$ is defined in Eq.~\eqref{M1-ys} and $\op M_\text{h}(y_s)$ is accounted terms of the third and forth order of expansion, defined as
\begin{align}\label{wwchv2}
    \op M_\text{h}(y_s)&\equiv \frac{1}{6}\left[ \op{c}^3(y^3_s\dt-3y_s)-2y_s\op{c}^\dagger\op{c}^2-y_s\op{c}\op{c}^\dagger\op{c}\right]
    +\frac{1}{24}\left[ \op{c}^4(y^4_s\dt^2-6y^2_s\dt+3)-(3\op{c}^\dagger\op{c}^3+2\op{c}\op{c}^\dagger\op{c}^2 + \op{c}^2\op{c}^\dagger\op{c})(y_s^2\dt-1)\right].
\end{align}

By computing the product of $\op M_2(y_s)$ similarly to Appendix~\ref{app-itomap} and introducing a dummy index $j$ for the infinitesimal record where $s= t+(j-1)\dt$, we can construct the high-order measurement operator as 
\begin{align}
\hat M_\text{W}(Y_t) &= \lim_{m\rightarrow \infty} \hat M_2(y_{t+(m-1)\dt})  \cdots \hat M_2(y_{t+\dt}) \hat M_2(y_{t}) \\
&= \lim_{m\rightarrow \infty} \prod_{j = 1}^{m} \left\{ \op M_1(y_j)+\left[\tfrac{1}{24}(\op{c}^\dagger\op{c})^2+\tfrac{1}{12}(\op{c}^\dagger)^2\op{c}^2 + \op M_\text{h}(y_j)\right]\dt^2 \right\},\\
&=\op 1 + \lim_{m\rightarrow \infty} \bigg\{ -\tfrac{1}{2}{m\choose m-1}\op c^\dagger\op c\dt+\op c\sum_j y_j\dt +\tfrac{1}{2}\op c^2\bigg[\sum_{j}y_j^2\dt^2-\dt{m\choose m-1}+\sum_{j \ne j'} y_j y_{j'}\dt^2\bigg]+\tfrac{1}{4}{m\choose m-2}(\op c^\dagger\op c)^2\dt^2 \nonumber\\
&-\tfrac{1}{2}\dt \sum_j {m-j\choose 1} y_j\dt\left[\op c\op c^\dagger\op c + c^\dagger\op c^2\right] +\tfrac{1}{4}\op c^3\left[\sum_{j}\dt y_j \left(\sum_{j'}y_{j'}^2\dt^2-\dt{m\choose m-1}\right)\right] +\frac{1}{24}{m\choose m-1}(\op{c}^\dagger\op{c})^2\dt^2 \nonumber\\
& +\frac{1}{12}{m\choose m-1}(\op{c}^\dagger)^2\op{c}^2\dt^2+\sum_j \op M_2(y_j) \dt^2 \bigg\},\label{Mwline3}\\
&=\hat{1}-\tfrac{1}{2}\hat{c}^\dagger\hat{c}\ddt+\hat{c}\ddt Y_t+\tfrac{1}{2}\hat{c}^2[Y_t^2\ddt^2-\ddt]+\tfrac{1}{8}\big(\hat{c}^\dagger\hat{c}\big)^2\ddt^2 -\tfrac{1}{4}\big(\hat{c}\hat{c}^\dagger\hat{c}+ \hat{c}^\dagger\hat{c}^2\big)\ddt^2Y_t,\label{derive-WWC}
\end{align}
where we have used the coarse-grained record, $Y_t$, as used in Eq.~\eqref{coarse-grain-weight-1}. The summation in the coefficient of $\op c\op c^\dagger\op c + c^\dagger\op c^2$ in Eq.~\eqref{Mwline3} can be evaluated as
\begin{align}
   \lim_{m\rightarrow \infty} \dt \sum_{j=1}^{m} (m-j)y_j\dt&=\int_t^{t+\ddt}[\ddt-(s-t)]y_s \dd s\equiv R_t.
\end{align}
We find that the integral $R_t=\frac{1}{2}\ddt^2 Y_t-Z_t$, where $Z_t$ is defined in Eq.~\eqref{YZdis}. This then leads to $-\tfrac{1}{2}[\op c\op c^\dagger\op c + c^\dagger\op c^2](\tfrac{1}{2}\ddt^2Y_t-Z_t)$. Since we consider the second order in $\ddt$ and we know that $\text{E}\{Z_t^2\} =\mathcal{O}(\ddt^3)$, we therefore only keep the lower-order coarse-gained record, $Y_t$, as appeared in the last term in Eq.~\eqref{derive-WWC}. Including $Z_t$ in the map might improve the accuracy of individual quantum trajectories, which will be investigated in future work. %our forthcoming paper~\cite{WWC2023}.

The other vanishing terms are a result of: (1) the property in Eq.~\eqref{sumys2}, which allows us to replace $\sum_j y_j^2\dt^2-\ddt =0$; and (2) the terms converge to zero as $m\rightarrow \infty$.

\section{Multi-variable integration}\label{moments-ZY}
To compute the moments, the record variables are discretized into $m=\ddt/\dt$ segments. Again, introducing the index $s= t+(j-1)\dt$, we have
\begin{subequations}\label{multi-inegrate} 
\begin{align}
\text{E}\{X_t^p \}&=\lim_{m \rightarrow \infty} \left(\prod_{j=1}^{m}\int \dd y_j \wp_{\rm{ost}}(y_j) \right)\left(\sum_{j=1}^{m}y_{j}\dt e^{-\gamma j\dt/2} \right)^p,\\
\text{E}\{Y_t^p Z_t^q \}&=\lim_{m \rightarrow \infty} \left(\prod_{j=1}^{m}\int \dd y_j \wp_{\rm{ost}}(y_j) \right) 
\left(\frac{1}{\ddt} \sum_{j=1}^{m}y_{j} \dt \right)^p \left(\sum_{j=1}^{m}y_{j} \dt [j\dt -(t+\tfrac{\ddt}{2})] \right)^q,
\end{align}
\end{subequations}
\end{widetext}
\lipsum[0]
where $p,q \in \{0,1, 2\}$. Following $\wp_\text{ost}(y_j)$ in Eq.~\eqref{ostprob-dt}, it is straightforward to evaluate the mean via \begin{align}
    \left(\prod_{j}\!\int \!\! \dd y_j \wp_{\rm{ost}}(y_j)\right) \sum_{j}y_{j}&=0.
\end{align}
For the first moment of each variable in Eqs.~\eqref{multi-inegrate}, the three variables therefore have zero mean, i.e., $\text{E}\{ X_t \}=\text{E}\{ Y_t \}=\text{E}\{ Z_t \}=0$.

Similarly, one can compute to the second moment via
\begin{align}\label{secmoment}
    \left(\prod_{j}\!\int \!\! \dd y_j \wp_{\rm{ost}}(y_j)\right) \sum_{j, j'}y_{j}y_{j'}&=\sum_{j, j'}\frac{\delta_{j, j'}}{\dt}.
\end{align}
Using the property in Eq.~\eqref{secmoment}, the resulting second moments are evaluated to the following integrals: 
\begin{align}
    \text{E}\{ X_t^2 \} &= \int_t^{t+\ddt}e^{-\gamma s}\dd s=\frac{2}{\gamma}e^{-\gamma(t+\ddt/2)}\sinh(\gamma\ddt/2),\nonumber\\
    \text{E}\{ Y_t^2 \} &=\frac{1}{\ddt^2}\int_t^{t+\ddt}\dd s=\frac{1}{\ddt},\nonumber\\
    \text{E}\{ Z_t^2 \} &=\int_t^{t+\ddt}[s-(t+\ddt/2)]^2\dd s=\ddt^3/12, \ \text{and}\nonumber\\
    \text{E}\{ Y_t Z_t \} &=\frac{1}{\ddt}\int_t^{t+\ddt}[s-(t+\ddt/2)]\dd s=0.
\end{align}

\blk

\section{Bayesian fluorescence map}\label{flu-bayesian}
Application of the Bayesian probability for the qubit fluorescence has been investigated in~\cite{Jordan2016}. By considering a composite system (the two-level system and the photon) and assuming its initial state is a product state given by
\begin{eqnarray}\label{eq-intdecay}
\ket{\psi_0}&=&(a\ket{\text{e}}+b\ket{\text{g}}) \ket{0},
\end{eqnarray}
where the qubit's initial state is a superposition state $a\ket{\text{e}}+b\ket{\text{g}}$, with $|a|^2+|b|^2=1$, and the bath's initial state is assumed to be a vacuum state. The qubit's decay can be treated phenomenologically, by considering the probability of the transition (excited state to ground state) for a time lenght $\ddt$. The probability of transition, emitting a single photon to the bath, given that the initial state of the qubit is $\ket{\text{e}}$ is $\wp(1|\text{e})=\gamma\ddt$, where $\gamma$ is the qubit's decay rate. This gives the probability of no decay as $\wp(0|\text{e}) = 1-\gamma\ddt$. The qubit's state then evolves from Eq.~\eqref{eq-intdecay} for the duration of time $\ddt$ to a final state given by
\begin{align}\label{eq-finspon}
\ket{\psi_f}=&\, a\sqrt{\wp(0|\text{e})}\ket{\text{e}}\ket{0}+a\sqrt{\wp(1|\text{e})}\ket{\text{g}}\ket{1}+b\ket{\text{g}}\ket{0} \nonumber \\
= &\, a\sqrt{1-\gamma\ddt}\ket{\text{e}}\ket{0}+a\sqrt{\gamma\ddt}\ket{\text{g}}\ket{1}+b\ket{\text{g}}\ket{0}.
\end{align}
From this initial and final state, we can construct the unitary operator for the composite system $\hat{U}_{t+\ddt,t}$ and derive the Kraus operator via $\ket{\psi_f}=\hat{K}_{\rm F, Bay}\ket{\psi_i}$.

Following Eq.~\eqref{MeasOperator}, the bath's initial state is the vacuum state, i.e., $|e_0 \ra = |0\ra$. The bath's final state will depend on the type of measurement considered. In this work, we consider the diffusive-type measurement, such as the homodyne detection.
%, which measures the quadrature observable, $\hat X_{\phi} = \hat a e^{-i \phi}+ \hat a^\dagger e^{+i \phi}$, of the bosonic bath. Here, $\hat a$ and $\hat a^\dagger$ are the annihilation and creation operators for the bath and $\phi$ is the phase of the homodyne measurement. The eigenstate of the quadrature observable can be given by $\hat X_{\phi} \ket{X} = X_{\phi}  \ket{X}$, where 
%\begin{align}\label{eq-eigenhomo}
%\la  X | 0\ra = \pi^{-1/4} e^{-X^2 /2}, \quad \la X | 1\ra = \pi^{-1/4}\sqrt{2}Xe^{-X^2 /2},
%\end{align}
%where we can see that these in Eq.~\eqref{eq-eigenhomo} are a wave of the ground and exited state of Harmonic oscillator. We can rescale the measurement result to our notation by $X = \sqrt{\ddt/2} X$. Using Eq.~\eqref{MeasOperator}, \eqref{eq-finspon}, and \eqref{eq-eigenhomo}, one can construct the Kraus operator for the continuous monitoring of the qubit's fluorescence,
For consistency of the notation to our work and projecting to the homodyne eigenstate, the measurement operator derived in Ref.~\cite{Jordan2016,Philippe2020} is given by
\begin{align}\label{MoFlu}
\hat{K}_{\rm F,Bay}(Y_t)&=\bigg(\frac{\ddt}{2\pi}\bigg)^{\frac{1}{4}}e^{-Y_t^2\ddt/4}\begin{pmatrix}\sqrt{1-\gamma\ddt}&0\\ \sqrt{\gamma}Y_t\ddt &1\end{pmatrix}.
\end{align}
The top-right element is simply the first order expansion of the exponential function of the top-right element in Eq.~\eqref{FluExact}, while the bottom-right element corresponds to the zeroth-order of expansion of $X_t$ defined in Eq.~\eqref{defX}.

\par The quantum Bayesian approaches are trace preserving (C3), which can be shown as
\begin{align}
    \int \dd Y_t \op K_{\rm{F,Bay}}^\dagger(Y_t)\op K_{\rm{F,Bay}}(Y_t)&=\hat{1}.
\end{align}
However, the map gives the average trajectory agreeing with the (B) Lindblad master equation only to the first order in $\ddt$. To see this, we expand $\op K_\text{F,Bay}(Y_t)$ to second order in $\ddt$. We find that it exactly coincides with $\op K_\text{G}(Y_t)$ as defined in Eqs.~\eqref{GWMO} and~\eqref{GWMO2} with $\op c=\sqrt{\gamma}\op\sigma_-$.

%------------------------------------------
% Reference
%------------------------------------------
\bibliographystyle{apsrev4-1}
\bibliography{RefSparque}

\end{document}